\documentclass[aps,prd,amssymb,eqsecnum,nofootinbib,floatfix, a4paper,twocolumn]{revtex4}

\usepackage{mathrsfs}
\usepackage{latexsym,amsmath,amsfonts,amssymb}
\usepackage{bm}
\usepackage{float}

\usepackage{graphicx}

\usepackage{xcolor}  

\allowdisplaybreaks

\newcommand{\be}{\begin{equation}}
\newcommand{\ee}{\end{equation}}
\newcommand{\bea}{\begin{eqnarray}}
\newcommand{\eea}{\end{eqnarray}}
\def\const{\mbox{const}}

\def\k{\kappa}
\def\lam{{\lambda}}

\def\d{\partial}
\def\l{\left(}
\def\r{\right)}

\def\t0{\tilde{0}}

\def \Yb{\overline{Y}}
\def \pb{\overline{\pi}}
\def \Lb{\overline{L}}
\def \hr{\hat{r}}
\def \hC{\overline{C}}

\def\del{{\nabla}}

\newcommand{\bg}{\begin{gather}}
\newcommand{\eg}{\end{gather}}
\newcommand{\bseq}{\begin{subequations}}
\newcommand{\eseq}{\end{subequations}}

\begin{document}

\title{Absence of a Vainshtein radius in torsion bigravity
}

\author{Vasilisa \surname{Nikiforova}}
 
\affiliation{Institut des Hautes Etudes Scientifiques, 
91440 Bures-sur-Yvette, France}

\date{\today}

\begin{abstract}
It was pointed out long ago by Vainshtein [Phys.\ Lett.\  {\bf 39B}, 393 (1972)] that the weak-field perturbation expansion
of generic theories (of the nonlinear Fierz-Pauli type) involving massive spin-2 excitations breaks down below a certain
distance around a material source (``Vainshtein radius''), scaling as some inverse power of the spin-2 mass $m_2$, {\it i.e.},
some positive power of the range $m_2^{-1}$. Here
we prove that this conclusion {\it does not apply} in a generalized Einstein-Cartan theory (called ``torsion bigravity'') whose
spectrum is made (like that of bimetric gravity) of a massless spin-2 excitation and a massive spin-2 one.  
Working within a static spherically symmetric ansatz, we  prove, by reformulating the field equations in terms of new variables,
that one can construct an all-order weak-field perturbative expansion where no denominators involving $m_2$ ever appear
in the region $r \ll m_2^{-1}$. In particular, we show how the formal large-range limit, $m_2 \to 0$, leads to a well-defined,
finite perturbation expansion, whose all-order structure is discussed in some detail.
\end{abstract}

\maketitle

\section{Introduction} \label{sec1}

Theories involving  massive spin-2 excitations raise several delicate issues. These issues arise both in massive gravity theories,
and in bimetric gravity theories. See Refs. \cite{deRham:2014zqa,Schmidt-May:2015vnx} for introductions to, respectively, 
massive gravity and bimetric gravity theories. The present work will study some of these issues within a new type of bigravity theory,
dubbed ``torsion bigravity", which has not yet been studied in detail. Torsion bigravity \cite{Damour:2019oru} is a geometric
theory, involving both massless spin-2 and massive spin-2 excitations, whose basic fields are a metric and an independent connection.
The massive spin-2 degrees of freedom are contained within the torsion of the independent connection.

The first general issue raised by the presence of massive spin-2 excitations is the so-called Boulware-Deser ghost \cite{Boulware:1973my}, 
namely,
a sixth degree of freedom appearing at the nonlinear level and having unbounded negative energy, and thus being pathological. 
For many years, it was thought that nonlinear massive gravity necessarily suffered from the presence of a Boulware-Deser ghost. 
The study of the decoupling limit \cite{ArkaniHamed:2002sp} allowed one to isolate the dangerous nonlinear couplings giving
rise to the Boulware-Deser ghost (see notably Refs. \cite{Creminelli:2005qk,Deffayet:2005ys,deRham:2010ik}).
Then it was found that by choosing special nonlinear mass terms one could eliminate this sixth ghostlike degree of freedom \cite{deRham:2010kj}. The latter de Rham-Gabadadze-Tolley ghost-free massive gravity theories were then generalized to ghost-free bimetric
gravity theories \cite{Hassan:2011zd} involving both a massless spin-2 excitation and a  massive spin-2 one. 

An important point for the present study is that, as shown in Ref. \cite{Babichev:2009us}, the presence or absence of a sixth,
ghostlike degree of freedom is visible in the simple setting of  static spherically symmetric solutions. More precisely, 
Ref. \cite{Babichev:2009us}  showed that the presence of the Boulware-Deser ghost was directly related to the order of differentiation,
in the field equations, of the function $\mu$ which describes the relation between the Schwarzschild radius (say $r$),
defined by the curved metric, and the Minkowski radius (say $r_\eta$), namely $r_\eta= r e^{- \mu(r)/2}$.
See the discussion in
section 4 of Ref. \cite{Babichev:2009us} where it is shown that the presence (in generic, ghostfull massive gravity theories)
of the second radial derivative  $\mu^{\prime\prime}$ in the quadratic part $Q(\mu)$ of the constraint 
coming from the Bianchi identity\footnote{In the notation of Ref. \cite{Damour:2002gp} the latter constraint reads
 $0=f_g(\lambda,\mu,\nu, \lambda^\prime,\mu^\prime,\nu^\prime,\mu^{\prime\prime},r)$ where $f_g \propto \del^\mu T_{\mu r}^g$.}
is  directly related to the higher-derivative nature of the scalar mode in the Goldstone picture \cite{ArkaniHamed:2002sp}.
As a consequence, the single equation (Eq. (4.4) in \cite{Babichev:2009us}) satisfied by $\mu$ is of the third differential
order, so that the general (exterior) solution of ghostfull massive gravity  contains {\it three integration constants.}\footnote{Our
counting of arbitrary constants here does not use any asymptotic boundary condition, {\it i.e.}, it  allows for exponentially growing
solutions at infinity.} By contrast, the general exterior solution of ghostfree massive gravity contains only  {\it two} arbitrary integration 
constants, {\it e.g.} \cite{Gruzinov:2011mm}, one of them corresponding to an exponentially growing solution. It was pointed out 
in Ref.\cite{Damour:2019oru} (and will be confirmed below) that the massive spin-2 sector of torsion bigravity is similar to ghostfree
massive gravity in that its general exterior spherically symmetric static solutions involve only two arbitrary constants, one of them
describing an exponentially growing solution. When considering both the massive and the massless spin-2 sectors, torsion
bigravity solutions involve (similarly to ghostfree bimetric gravity) three arbitrary integration constants, the third one corresponding
to the Einsteinlike massless spin-2 sector, and describing a Schwarzschildlike mass.

The second issue  raised by the presence of massive spin-2 excitations is the question of the so-called Vainshtein mechanism, or Vainshtein screening. When one solves the equations of motion in massive gravity within the framework of perturbation theory, already at the linear level there appears the term $(m_2 r)^2$ as a denominator. Here $m_2$ denotes the mass of the massive spin-2 excitation\footnote{In our 
torsion bigravity discussion below, we will use the notation $\kappa$ for $m_2$.}, with units of inverse length, such that $m_2^{-1}$ defines
the range of the massive spin-2 interaction.
In the higher orders of perturbation theory, there appear denominators with increasing powers $(m_2 r)^n$ (see, {\it e.g.} \cite{Damour:2002gp}). Vainshtein \cite{Vainshtein:1972sx} pointed out that the presence of such denominators  limits the domain of validity
of perturbation theory to distances $r \gtrsim r_V$, where $r_V$ is the so-called Vainshtein radius. In the case of generic,
ghostfull, massive gravity $r_V$ is of order 
\be 
r_V = \left( \frac{GM}{m_2^4} \right)^{1/5}\,,
\ee
where $M$ denotes the mass of the star. When considering a small $m_2$, i.e. a large range $m_2^{-1}$ (say of cosmological size),
$r_V$ is typically much larger than the length scales where gravity has been accurately checked to be in close
agreement with General Relativity (GR). 
Vainshtein has argued \cite{Vainshtein:1972sx} (see also \cite{Deffayet:2001uk})
that in the region  $r \lesssim r_V$
there existed an alternative series expansion, involving positive powers of $\frac{r}{r_V}$, and that these two different expansions 
(the first one involving positive powers of  $\frac{r_V}{r}$, and valid for $r \gg r_V$, and the other one valid for $r\ll r_V$) 
 merge at  distances $r  \sim r_V$. 
This claim of  Vainshtein was proven to hold true in the case of static spherically symmetric solutions
by Babichev, Deffayet and Ziour \cite{Babichev:2009jt,Babichev:2010jd}. The latter references showed 
the existence of a global solution matching the two just-mentioned expansions. See {\it e.g.} Ref. \cite{Babichev:2013usa}
for a discussion of the cases
where the Vainshtein mechanism has been shown to hold.

Finally, the third problem that massive gravity faces  is the van-Dam-Veltman-Zakharov (vDVZ)  discontinuity \cite{  vanDam:1970vg, Zakharov:1970cc}. The vDVZ  discontinuity is the fact that the  massless limit $m_2\to0$ of light deflection in massive gravity 
differs by a factor $\frac34$ from light deflection in GR. 
 The origin of this finite difference is the coupling to the trace of energy-momentum tensor, in massive gravity, of an additional scalar degree of freedom. This additional coupling follows from the usual five degrees of freedom of a massive spin-2 field, as present
 in linear Fierz-Pauli theory.

The vDVZ phenomenon is not theoretically problematic {\it per se}, but it is phenomenologically problematic in that it seems to require,
when considering a theory involving {\it both} massless and massive spin-2 fields, that the coupling of the massive spin-2 field
be tuned to a small value, so as to be consistent with experimental tests of GR. This phenomenological issue is drastically
changed when the  Vainshtein mechanism is at work. Indeed, in that case, if $m_2$ is very small (say of cosmological magnitude),
 usual perturbation theory is invalid, so that one cannot use the prediction of light bending based on weak-field perturbation theory
 (see, for example, \cite{Deffayet:2001uk}). The Vainshtein mechanism (when it is at work) modifies the predictions of the theory at $r \lesssim r_V$,
 and effectively screens the effects of massive gravity to recover the predictions of GR,
 thereby matching the Solar system observations.

The aim of the present paper is to study whether an analog of the Vainshtein mechanism is present within torsion bigravity \cite{Damour:2019oru}. Torsion bigravity is a geometric theory which generalizes the Einstein-Cartan theory by having, besides a dynamical metric,
a propagating torsion. It was defined in Ref. \cite{Damour:2019oru} as a special case of the multi-parameter class of
ghost-free and tachyon-free theories with propagating torsion introduced in 
Refs. \cite{  Sezgin:1979zf, Sezgin:1981xs, Hayashi:1979wj, Hayashi:1980av, Hayashi:1980ir, Hayashi:1980qp}. 
The spectrum (around Minkowski space) of torsion bigravity comprises only massless spin-two excitations  together with massive 
spin-two ones. The previous paper \cite{Damour:2019oru} began the study of static spherically symmetric solutions in this model.
Some remarkable features of this model were found. 

First, as already mentioned, the counting of the number of arbitrary integration constants in the general
exterior  static spherically symmetric was found to be equal to three, which is the same number of 
of  integration constants needed to describe general exterior solutions  in {\it ghost-free} bimetric gravity theories \cite{Volkov:2012wp}.
Taking into account the  situation in generic massive gravity theories \cite{Babichev:2009us}, this is an indication
 of the absence of Boulware-Deser ghost in this model.  

Second, as torsion bigravity contains a massive spin-2 excitation (with mass denoted henceforth as $\k =m_2$), it is {\it a priori}
expected that its weak-field  perturbation theory will involve (similarly to all known ghostfull or ghostfree nonlinear
Fierz-Pauli models) denominators proportional to  powers of the mass, thereby signalling the breakdown of perturbation theory
at some Vainshteinlike radius.  However, it was found
in Ref. \cite{Damour:2019oru} that: (i) at the linear level of perturbation theory, no denominators appeared (contrary to what happens
even in the linear Fierz-Pauli model); and, (ii) at the quadratic level ($O(G^2)$) of perturbation theory, there happened remarkable 
cancellations between various terms of order $\k^{-2}$ in the field equations leading to a final second-order solution which did not have
any singularity in the massless limit $\k \to 0$. As Ref. \cite{Damour:2019oru} could not decipher any deep reason behind the
cancellations between the second-order   $O(\k^{-2})$ terms, it left undecided the issue of whether such cancellations would occur
to higher orders, or would stop occurring at the third order.

The main result of the present paper will be to present a simple explanation for the occurrence of the cancellations found in 
Ref. \cite{Damour:2019oru}, and to show  that such cancellations actually occur {\it at
all orders} of perturbation theory. This will be done by first reformulating the field equations in terms of new variables, and
showing that the corresponding transformed field equations contain only positive powers of $\k^2$ (while the original field
equations did involve some  $\k^{-2}$ factors). Then, using these reformulated field equations (involving new variables),
we will  show that they can be globally solved (both in the source and in the exterior domain $r \lesssim \k^{-1}$)
without ever introducing inverse powers of $\k$ in the solution. In other words, the small-mass limit of torsion bigravity
does not exhibit any Vainshteinlike radius (scaling with some inverse power of $\k$) indicating a breakdown of perturbation theory.
This means that perturbation theory holds in torsion bigravity even at  small distances from (or inside) the source.  

This absence of any Vainshtein radius in torsion bigravity is a remarkable theoretical fact which, however, has a somewhat
unpleasing phenomenological consequence. Indeed, the vDVZ discontinuity is still present in torsion bigravity (because the
massive spin-2 piece of the solution couples to the energy-momentum tensor in a different way than the massless spin-2 one).
The absence of any Vainshtein radius and, thus, the absence of any putative Vainshtein mechanism, makes it impossible to screen 
the vDVZ discontinuity by a nonlinear modification at small distances, as it was the case in massive gravity and in
(massive) bimetric gravity. This implies that we must constrain the coupling constant linked to the massive spin-two exchange
to a small enough value, so as to be consistent with the current (GR-compatible) experimental limits on post-Newtonian
gravity  (see Sec.~X of \cite{Damour:2019oru} for details). 


\section{Formalism and definitions}

We follow the notation of Ref. \cite{Damour:2019oru}. Let us only recall some basic notational features.
We work with two independent fields:
the vierbein ${e_i}^{\mu}$ (with inverse ${e^i}_{\mu}$, ${e_i}^\mu{e^j}_\mu =\delta^j_i$), and the SO(3,1) connection  ${A^i}_{ j \mu}$ which is constrained to be {\it metric preserving} (i.e. satisfying the condition of antisymmetry $A_{i j \mu}= - A_{j i \mu}$, where $A_{i j \mu}\equiv \eta_{is} {A^s}_{ j \mu}$).
We use Greek letters $\mu, \nu, ... =0,1,2,3$ to denote spacetime indices, which are linked to a coordinate system $x^{\mu}$ and moved by the coordinate-system metric
$g_{\mu \nu} \equiv \eta_{ij} {e^i}_\mu {e^j}_\nu$. Latin indices $i, j, k, ... =0,1,2,3$ are used to denote the Lorentz-frame indices linked to the vierbein ${e_i}^{\mu}$; they are moved by the Minkowski metric $\eta_{ij}$ . When there is a risk of ambiguity we add a hat marking frame index, for example, ${e_{\hat{i}}}^{\mu}$. We will always try to place the frame index before the coordinate one (``frame first" rule). We use a mostly plus signature. 

From the (inverse) vierbein ${e^i}_{\mu}$, we can derive the Levi-Civita connection $ {\omega^i}_{j\mu}(e)$. The difference between 
our general connection ${A^i}_{ j \mu}$ and the Levi-Civita connection $ {\omega^i}_{j\mu}(e)$ defines the 
{\it contorsion tensor}  ${K^i}_{ j \mu} $ 
\be \label{KvsAe}
{K^i}_{ j \mu} \equiv {A^i}_{ j \mu} - {\omega^i}_{j\mu}(e).
\ee
The frame components ${K^i}_{ j k} \equiv {e_k}^\mu {K^i}_{ j \mu}$
of the contorsion tensor are related to the frame components ${T^i}_{[jk]}= - {T^i}_{[kj]}$ of the torsion tensor as follows
\bea
K_{ijk}&=& \frac12 (  T_{i [jk]}+ T_{j [ki]} - T_{k[ij]}) \,, \nonumber\\
T_{i [jk]} &=&  K_{ijk}- K_{ikj}.
\eea

Among the various possible Lagrangians defining the family of ghost-free and tachyon-free dynamical torsion theories (see \cite{  Sezgin:1979zf, Sezgin:1981xs, Hayashi:1979wj, Hayashi:1980av, Hayashi:1980ir, Hayashi:1980qp} for a discussion of the possible actions
defining this family, and the corresponding field contents), the Lagrangian of torsion bigravity selects the models containing only one massless spin-2 excitation, and one massive spin-2 one. It has {\it four parameters}, and its action reads
\be \label{Stot}
S_{\rm total} = S_{\rm TBG}[{e^i}_\mu, A_{i j \mu}]+ S_{\rm matter} \,,
\ee
where the essential torsion bigravity part $ S_{\rm TBG}$ is 
\be \label{STBG}
S_{\rm TBG}[{e^i}_\mu, A_{i j \mu}] =\int d^4 x \, \sqrt{g} \,  L_{\rm TBG}[e, \d e, \d^2 e, A, \d A]\,,
\ee
with $\ \sqrt{g} \equiv  \sqrt{ - \det g_{\mu \nu}} \equiv \det {e^i}_\mu$, and
\bea  \label{lagrangian}
L_{\rm TBG} &=& c_R \, R[e, \d e, \d^2 e] +c_F \, F[e, A, \d A] \\ \nonumber
&+& c_{F^2}\left( F_{(ij)} F^{(ij)} - \frac13 F^2 \right)
 + c_{34} F_{[ij]} F^{[ij]}  \; . 
 \eea
Here we use the letter $R$ to denote the various Riemannian curvature structures derived from the Levi-Civita connection $ {\omega^i}_{j\mu}(e)$ , such as the Riemannian curvature tensor ${R^i}_{jkl} \equiv  {R^i}_{ j \mu \nu} {e_k}^\mu {e_l}^\nu$, the Ricci tensor 
 $R_{ij}={R^k}_{ikj}$ and the curvature scalar $R=\eta^{ij}R_{ij}$. The objects denoted by the letter $F$ are the various Yang-Mills curvature structures derived from the connection ${A^i}_{ j \mu}$, such as the curvature tensor ${F^i}_{jkl} \equiv  {F^i}_{ j \mu \nu} {e_k}^\mu {e_l}^\nu$, the Ricci tensor 
 $F_{ij}={F^k}_{ikj}$ and the curvature scalar $F=\eta^{ij}F_{ij}$. Looking at the structure of the Lagrangian \eqref{lagrangian} with respect to derivatives $\d e$ and $\d A$ one can see that the equations of motion will be at most second order in the derivatives of $e$ and $A$. 
 
 The three parameters $c_R$, $c_F$ and $c_{F^2}$ have simple physical meanings. Namely, $c_R$ and $c_F$ define the coupling constants $G_0$ and $G_m$ linked to massless spin-2 exchange and massive spin-2 exchange, respectively. More precisely, the sum of these two constants yields the usual massless gravitational coupling according to
 \be
 c_R + c_F \equiv  \lam = \frac{1}{16 \pi G_0} \label{cR}\,,
 \ee
 while the ratio $c_F/c_R$ measures the ratio of the two couplings to matter
 \be
 \frac{c_F}{c_R} \equiv  \eta = \frac{3}{4} \frac{G_m}{G_0} \label{cF}\,.
 \ee
 The parameter $c_{F^2}$ is linked to the mass of of the massive spin-2 excitation which we denote as $\k$
\be
c_{F^2} = \frac{\eta \, \lam}{\k^2} = \frac{ c_F (1 + \frac{c_F}{c_R})}{\k^2} \,. \label{cF2}
\ee
The fourth parameter $c_{34}$ has no evident physical meaning, but it does not enter into the discussion of
spherically symmetric solutions.

\subsection{Static spherically symmetric case}
In the present paper we continue to investigate static spherically symmetric solutions in torsion bigravity. This means that the solutions we study satisfy the following three conditions:
(i) time-reversal invariance; (ii) SO(3) invariance; and (iii) parity invariance. The parameter $c_{34}$ does not contribute to the discussion of the spherically symmetric case, because the conditions (i)-(iii) imply that the antisymmetric part of the tensor $F_{[ij]}$ vanishes.

It is useful to work in Schwarzshildlike coordinates $x^{\mu}:\,\{x^0\equiv t, x^1\equiv r, x^2 \equiv \theta, x^3 \equiv \phi\}$, so that we write the metric as follows
\be \label{ds2}
ds^2=-e^{2\Phi}dt^2 + e^{2\Lambda}dr^2 + r^2\left( d\theta^2+\sin^2\theta\, d\phi^2 \right) \; .
\ee
We use as (inverse) orthonormal frame ${e^{\hat{i}}}_{\mu}$ associated to our coordinate system $x^{\mu}$
the one whose nonzero components  are 
\be \label{frame}
{e^{\hat{0}}}_0=e^{\Phi} \;, \quad {e^{\hat{1}}}_1=e^{\Lambda} \;, \quad {e^{\hat{2}}}_2=r  \;, \quad {e^{\hat{3}}}_3= r \sin\theta \;.
\ee

As explained in Ref. \cite{Damour:2019oru}, our assumptions (i)-(ii)-(iii) above imply that the SO(3,1) connection involves
only two functions of $r$, entering the following nonzero components of ${A^i}_{ j \mu}$:
\begin{eqnarray} \label{VW}
V(r)&=&{A^{\hat{1}}}_{\hat{0}\hat{0}}=+{A^{\hat{0}}}_{\hat{1}\hat{0}} \,,  \nonumber \\
W(r)&=&{A^{\hat{1}}}_{\hat{2}\hat{2}}={A^{\hat{1}}}_{ \hat{3}\hat{3}}=-{A^{\hat{2}}}_{\hat{1}\hat{2}} = -{A^{\hat{3}}}_{\hat{1}\hat{3}}\;.
\end{eqnarray}
Thus, our basic field variables are $\Phi(r)$ and $\Lambda(r)$ from Eq. \eqref{ds2} and $V(r)$ and $W(r)$ from Eq. \eqref{VW}.

From the vierbein \eqref{frame} one can compute the Levi-Civita connection $ {\omega^i}_{j\mu}(e)$, and then the contorsion from \eqref{KvsAe}. Its nonzero components read
\bea \label{contorsion}
{K^{\hat{1}}}_{\hat{0}\hat{0}}&=& {K^{\hat{0}}}_{\hat{1}\hat{0}}=V -  e^{-\Lambda}\Phi^{\prime} \,,\nonumber \\
{K^{\hat{1}}}_{\hat{2}\hat{2}}&=& {K^{\hat{1}}}_{\hat{3}\hat{3}}=W +  r^{-1}e^{-\Lambda} \,.
\eea
Here and below we use a prime to denote a radial derivative $\d_r$.

Armed with this knowledge, we are ready to look more precisely at the structure of the Lagrangian.

\section{The original Lagrangian and its reformulation}

As written in Eq. \eqref{Stot},   the total action is a sum of the field action and the matter action.
 The matter action  is such that the energy-momentum tensor $T^{\mu\nu}$ arises through its variation  with respect to the metric:
 \be
\delta S_{\rm{matter}}=\int\delta (\sqrt{g} L_m) d^4x  = \frac{1}{2}\int\sqrt{g}T^{\mu\nu}\delta g_{\mu\nu}d^4x \,.
\ee
The field action is defined by the Lagrangian \eqref{lagrangian} (with a vanishing $c_{34}$ contribution). It consists, in the static spherically symmetric case, of three 
different contributions
\be 
S_{\rm{TBG}}=S_R+S_F+S_{F^2} =\int d^4x  \sqrt{g}\left\{ L_R+L_F+L_{F^2} \right\} \label{SField}\,.
\ee
with
\bea
L_R&=&c_R R[g] \;, \nonumber \\
L_F &=& c_F F[g, A] \;, \nonumber \\
 L_{F^2}&=& c_{F^2}\left(F_{(ij)}^2 - \frac{1}{3}F^2\right)  \,.
\eea
Here, according to \eqref{ds2},
\be
d^4x  \sqrt{g}=  dt \, ( w(r) \,  dr )\, (\sin \theta \, d \theta \,d\phi) \,,
\ee
where
\be
w(r) \equiv r^2e^{\Phi+\Lambda}\,.
\ee

In a previous paper (see Eq. (4.19) in \cite{Damour:2019oru}), we derived an explicit expression of the action \eqref{SField}
\be S_{\rm TBG}=\int w(L_R+L_F+L_{F^2})dt \,dr (\sin{\theta} \, d\theta \,d\phi) \,. \label{L3Parts}\ee
To present this explicit expression, it is useful to introduce a shorthand notation for the  following covariantlike derivatives  
of the functions $V$ and $W$:
\begin{eqnarray}
\nabla V&\equiv& e^{-\Phi-\Lambda}(e^{\Phi}V)^{\prime}=e^{-\Lambda}\left(V^{\prime} + \Phi^{\prime} V\right) \,, \\
\nabla W &\equiv& e^{-\Lambda}\frac{(rW)^{\prime}}{r}= e^{-\Lambda}\left( W^{\prime} + \frac{W}{r}\right) \;.
\end{eqnarray}
We also use
\be
W^2_{-} \equiv W^2 - \frac1{r^2} \,.
\ee

With this notation, we found that the various contributions to the action \eqref{L3Parts} take the following form (after some integration by parts)
\be
w \,L_R=2c_R e^{\Phi+\Lambda}\frac{d}{dr} \left[ r(1-e^{-2\Lambda}) \right] + \frac{d}{dr}\left(c_R Q(r) \right)\,,\label{LR} 
\ee
where
\be
Q(r)=  -2r^2e^{\Phi-\Lambda} \Phi^{\prime} \; ;
\ee
\begin{eqnarray}
w \,L_F&= &c_F r^2 e^{\Phi+\Lambda}(4\nabla W - 2\nabla V + 4V W - 2W^2_{-}) \,, \label{LF}\nonumber \\
\end{eqnarray}
\begin{eqnarray}
\frac{3}{2c_{F^2}}w \,L_{F^2}&= &r^2e^{\Phi+\Lambda}\left\{  (\nabla V+ \nabla W)^2  \right.\nonumber \\
&&  + 2\nabla V(VW-2W^2_{-})+2 \nabla W(-5VW + W^2_{-}) \nonumber \\
& &\left. +(VW+W^2_{-})^2\right\}  \;. \label{LF21}
\end{eqnarray}

From these expressions  we see that the only part in the action containing the {\it square} of derivatives of $V^{\prime}$ and $W^{\prime}$ 
is  \eqref{LF21}, which reads
\bea
w \,L_{F^2} &=&  \frac{2c_{F^2}}{3} r^2e^{\Phi+\Lambda}\left\{ (\nabla V+ \nabla W)^2+...\right\}\nonumber \\
 &=& \frac{2\eta\lambda}{3\k^2} r^2e^{\Phi-\Lambda}\left\{ (V^{\prime}+W^{\prime}+...)^2 +...\right\} \,. \nonumber \\
 \eea
 After taking a Lagrange variation, this part will give second derivatives of $V$ and $W$, namely 
\bea 
\frac{\delta L_{F^2}}{\delta V} &=&-\frac{4\eta\lambda}{3\k^2}r^2e^{\Phi-\Lambda}(V^{\prime\prime} + W^{\prime\prime})+... \\
\frac{\delta L_{F^2}}{\delta W} &=&-\frac{4\eta\lambda}{3\k^2}r^2e^{\Phi-\Lambda}(V^{\prime\prime} + W^{\prime\prime})+...
\eea
Thus, the only second derivatives appearing in the equations of motion is the combination $V^{\prime\prime} + W^{\prime\prime}$. 
To minimize the appearance of second derivatives in the equations of motion, it is convenient to introduce the new variable,
\be
Y \equiv V+ W \;.
\ee
One can see from \eqref{contorsion} that in the flat limit, when $\Lambda\to0$ and ${K^{i}}_{jk} \to 0$, we have $W\to -r^{-1}$. It is then 
convenient to  replace $Y$ by the following new variable which is regular when $r\to0$
\be
\Yb \equiv Y+\frac{1}{r}=  V + W + \frac{1}{r} \,. \label{Ybdef}
\ee
After all these changes, the term in $L_{F^2}$ creating second-order derivatives reads
\be
w \,L_{F^2}=\frac{2\eta\lambda}{3\k^2} r^2e^{\Phi-\Lambda}\left(\Yb^{\prime}\right)^2 + \cdots
\ee

Since this is the only term containing the square of a derivative, it is convenient to add a so-called double-zero term to the action, 
so as to transform the action into one which is linear in derivatives, and thereby lead to first-order equations of motion.
 The general idea, when starting from  a Lagrangian of the form
$$ L_{\rm toy}(\dot{q},q)= \dot{q}^2+...$$
is to add to this lagrangian a term of the form
$$ \Delta L_{\rm toy}(\dot{q},q,\pi) = -(\dot{q} + f_1(q) - \pi)^2\,. $$
This compensates the $\dot{q}^2$ contribution
$$ L_{\rm toy}(\dot{q},q)+ \Delta L_{\rm toy}(\dot{q},q,\pi) =2 \dot{q} \pi - 2 \dot{q}f_1(q) + f_2(q,\pi) $$
so that the Lagrangian becomes linear in derivatives, and  the corresponding equations obtained from it are first-order in derivatives. 
This reduction to first order 
 is  achieved by the augmentation of the number of variables (and equations), namely by having replaced $\dot{q} +  f_1(q)$
 by the new momentumlike variable $\pi$.

In Ref. \cite{Damour:2019oru}, we had added a double-zero term of the form
\be
w \,\Delta L_{F^2} \equiv -  \frac{2\eta\lambda}{3\k^2} r^2e^{\Phi+\Lambda} \left(\nabla V+ \nabla W -\pi \right)^2 \,,\label{pintro}
\ee
essentially corresponding to replacing $\nabla V+ \nabla W$ by the momentumlike variable $\pi$.
This led to a modified  contribution \eqref{LF21}  of the form
\begin{eqnarray}
w \, {L}_{F^2}^{\rm{mod}} &=&  \frac{2}{3}c_{F^2}r^2e^{\Phi+\Lambda}\left\{ 2\pi (\nabla V+ \nabla W) \right.\nonumber \\
& & \left.  - \pi^2 + 2\nabla V(VW-2W^2_{-})  \right. \nonumber \\
 & & \left. +2 \nabla W(-5VW + W^2_{-})+(VW+W^2_{-})^2\right\}  \;. \label{LF2new} \nonumber \\
\end{eqnarray}

The Lagrangian $L=L_R+L_F+L_{F^2}^{\rm{mod}}$ with $L_R$, $L_F$ and ${L}_{F^2}^{\rm{mod}}$ given by \eqref{LR}--\eqref{LF}, \eqref{LF2new} achieves the purpose of leading to first-order equations of motion.
However, similarly to the original, unmodified second-order action \eqref{SField}, it has the feature
of explicitly containing  $c_{F^2} \propto \k^{-2}$, see Eq. \eqref{cF2}. 
Therefore the equations of motions for $\Phi, \Lambda, V, W$ and $\pi$ explicitly contain  factors $\k^{-2}$,
which are singular in the massless limit $\k \to 0$. 

Let us now show how one can introduce a new variable $\bar{\pi}$, different from $\pi$, which leads to a new first-order
Lagrangian containing only factors $\k^2$ instead of $\k^{-2}$.

First, let us prove that the action contribution $L_{F^2}$ can be compactly rewritten as
\begin{eqnarray}
w \,L_{F^2}&=&\frac{2}{3}c_{F^2}r^2e^{\phi+\Lambda}(\nabla V+\nabla W +V W +W_{-}^2)^2 \nonumber \\
&&+ \,\text{total derivative}\,. \label{SimpStr}
\end{eqnarray}
Indeed
\begin{eqnarray}
\frac{3}{2c_{F^2}}w \,L_{F^2}&= &r^2e^{\Phi+\Lambda}\left\{ (\nabla V+\nabla W +V W +W_{-}^2)^2+\Delta  \right\} \nonumber \,,
\end{eqnarray}
where
\begin{eqnarray}
\Delta&=&2\nabla V(V W-2W_{-}^2)+2\nabla W(-5VW+W_{-}^2) \nonumber \\
&&- 2(\nabla V+ \nabla W)(VW+W_{-}^2 ) \nonumber\\
&&= -6\nabla V W_{-}^2 - 12 \nabla W V W \,, 
\end{eqnarray}
so that
\begin{eqnarray} 
&&r^2 e^{\Phi+\Lambda}\Delta  = -6(e^{\phi}V)^\prime(W^2r^2-1)-12e^{\phi}VrW(rW)^\prime \nonumber \\
&& = -6\frac{d}{dr}\left[ e^{\phi}V(W^2r^2-1) \right] \,.
\end{eqnarray}
This exhibits a remarkably simple structure \eqref{SimpStr} for $w\, L_{F^2}$. 

It is then natural to add another double-zero term, instead of \eqref{pintro}, namely
\be
-\frac{2}{3}c_{F^2}r^2e^{\phi+\Lambda}(\nabla V+\nabla W +V W +W_{-}^2-c_{\pi}\bar{\pi})^2 \;,
\ee
where $c_{\pi}$ is a constant.
This term  modifies the structure of $L_{F^2}$ into
\begin{eqnarray}
w \,L_{F^2}^{\rm{new}} &=&  \frac{4}{3}c_{\pi}c_{F^2}r^2e^{\Phi+\Lambda}\pb(\nabla V+\nabla W +V W  \nonumber \\
&&  +W_{-}^2) -  \frac{2}{3}c_{\pi}^2c_{F^2}r^2e^{\Phi+\Lambda}\pb^2 \,. \label{Lnew}
\end{eqnarray}
The $O(\k^{-2})$ coefficient $c_{F^2}=\eta \lambda /\k^2$ enters \eqref{Lnew} only together with a factor $c_{\pi}$ or  $c_{\pi}^2$.  We can therefore eliminate the explicit presence of $\k^{-2}$ factors  in the action  by choosing, for instance, the following
value of the coefficient $c_{\pi}$:
\be
c_{\pi}\equiv\kappa^2 \;.
\ee
Then the prefactors in \eqref{Lnew} are 
\be
c_{\pi}c_{F^2}=\eta \lambda \, , \quad c_{\pi}^2c_{F^2}=\k^2 \eta \lambda \,.
\ee
The important point is that all the (first-order) equations of motion derived from $w \,L_{F^2}^{\rm{new}}$ involve 
no factors $\k^2$ in front of derivatives, and  some factors $\k^2$ in non-derivative terms.
For instance, the variation with respect to  $\pb$ yields the equation
\be
\nabla V+\nabla W +V W +W_{-}^2=c_{\pi}\bar{\pi}=\k^2\pb \,. \label{defPb}
\ee

One ends up with a new Lagrangian (given by formulas \eqref{L3Parts}, \eqref{LR}, \eqref{LF}, \eqref{Lnew}, with \eqref{cR}--\eqref{cF2}) as follows
\bea
&&w\, L_{TBG}= w\,L_R+w\,L_F+w \,L_{F^2}^{\rm{new}} \nonumber \\
&&=2\frac{\lam}{1+\eta} e^{\Phi+\Lambda}\frac{d}{dr} \left[ r(1-e^{-2\Lambda}) \right]  \nonumber \\
&& +\frac{\eta \lam}{1+\eta} r^2 e^{\Phi+\Lambda}(4\nabla W - 2\nabla V + 4V W - 2W^2_{-}) \nonumber \\
&&+   \frac{4}{3}\eta\lam r^2e^{\Phi+\Lambda}\pb(\nabla V+\nabla W +V W  \nonumber \\
&&  +W_{-}^2) -  \frac{2}{3}\eta\lam \k^2 r^2e^{\Phi+\Lambda}\pb^2 + \, \text{tot. der.}  \label{LagrNew}
\eea

The important fact for our purpose is that this Lagrangian does not contain any inverse powers of $\k$, and actually contains
a $\k^2$ factor only in the last algebraic term
\be
 -  \frac{2}{3}\eta\lam \k^2 r^2e^{\Phi+\Lambda}\pb^2 \,.
\ee
 As a consequence, the equations of motion obtained from the new lagrangian $L=L_R+L_F+L_{F^2}^{\rm new}$ contain no inverse powers of $\k$. This explains the remarkable cancellations we found at first and second perturbative orders in Ref. \cite{Damour:2019oru}.
 Actually, it is because we had also found similar cancellations at the third, fourth and fifth perturbative orders 
 when solving the equations of motion obtained from the old lagrangian $L=L_R+L_F+{L}_{F^2}^{\rm mod}$ that we looked
 for such a simple explanation. 
 
 Note that when $\k \to 0$, $\pb$ becomes a Lagrange multiplier, i.e. it appears linearly in the action and thus enforces the constraint
\be
\nabla V+\nabla W +V W +W_{-}^2 \approx 0 \quad \text{when} \quad \k=0\,.
\ee
However, $\pb$ enters other equations of motion ({\it e.g.} $\delta L/\delta V$  involves $\pb^{\prime}$), so that the Lagrange 
multiplier $\pb$ must be kept.

The fact that the new Lagrangian contains $\k^2$ instead of $\k^{-2}$ does not, by itself,
automatically prove that there exist a smooth limit of this model with $\k \to 0$. Indeed, let us recall the (apparently) similar case of  spherically symmetric solutions in  massive gravity (see, {\it e.g.}, the analysis of Damour et al \cite{Damour:2002gp}). There one has a Lagrangian which contains only the square $m_2^2 = \k^2$ of the mass,  and $\k^2$ appears only in factor of the algebraic mass term.
Nevertheless, the equations of motion imply several constraints whose solution necessitate to divide by  $\k^2$.
This introduces negative powers of $\k$ in the solution and, as recalled in the Introduction, renders the perturbative solution invalid for distances $r$ smaller than a certain scale (Vainshtein radius).
In the next section we are going to show that this does not happen in our case. Namely, we shall explicitly check
that all the equations of motion can be resolved with respect to the derivatives, without introducing inverse powers of $\k$. 

\section{Equations of motion}
We work (similarly to the previous paper \cite{Damour:2019oru}, modulo the replacement $\pi \to \pb$) 
with the set of  variables $\{F\,, \; L\,,\; V\,,\; \Yb\,,\; \pb \}$, where
\bea \label{defs}
L &\equiv&  e^{\Lambda} \,,\\
F &\equiv& \Phi^{\prime} \,.
\eea
Minkowski space with zero torsion is a solution of the model under consideration. We call it the {\it flat limit}, and, according to \eqref{ds2} and \eqref{contorsion}, our field variables take the following values in the flat limit 
\bea
L &\to 1  \, , \quad F\,,\;V\,,\; \Yb\,,\;\pb  &\to 0 \, . \label{flatlim}
\eea

Varying the Lagrangian \eqref{LagrNew} with respect to the five independent variables $\{ \Lambda\,,\Phi \,, \, V\,,\, W\,,\, \pb \}$ (it was more convenient to vary with respect to $W$ than to $\Yb$ when we derived equations of motion) 
yield five equations of motion. Among these five equations, there is one algebraic equation 
\be
E_{\Lambda}\equiv e^{-\Phi}\frac{\delta L^{\rm new}}{\delta \Lambda}=E_{\Lambda}(L,\,F, \,V,\,\Yb,\,\pb,\,\k^2)
\ee
and four first-order differential equations (we multiply each equation by $e^{-\Phi}$ in order to get rid of the variable $\Phi$ and to leave only $F$)
\bea
E_{\pb}&&\equiv e^{-\Phi}\frac{\delta L^{\rm new}}{\delta \pb}=E_{\pb}(\Yb^{\prime},\,L,\,F, \,V,\,\Yb,\,\pb,\,\k^2) \\
E_{V}&&\equiv e^{-\Phi}\frac{\delta L^{\rm new}}{\delta V}=E_{V}(\pb^{\prime},\,L, \,V,\,\Yb,\,\pb) \\
E_{W}&&\equiv e^{-\Phi}\frac{\delta L^{\rm new}}{\delta W}=E_{W}(\pb^{\prime},\,L,\,F, \,V,\,\Yb,\,\pb) \\
E_{\Phi}&&\equiv e^{-\Phi}\frac{\delta L^{\rm new}}{\delta \Phi}=E_{\Phi}(\pb^{\prime},\,\Yb^{\prime},\,V^{\prime},\,L^{\prime},\,L, \,V\,,\Yb,\,\pb,\,\k^2) \,.\nonumber  \\
\eea
We describe the material source by the following (perfect fluid) macroscopic energy-momentum tensor,
\be
T^{\mu\nu}=\left[ e(r)+P(r)\right]u^{\mu}u^{\nu}+P(r)g^{\mu\nu}\,,
\ee
where $u^{\mu}$ are the components of the 4-velocity. The components of the energy-momentum tensor for static spherically symmetric configurations are
\begin{eqnarray}
\label{MacroEMT}
T^{00}&=&\left[ e(r)+P(r)\right] e^{-2\Phi}-P(r)e^{-2\Phi} = e(r)e^{-2\Phi} \nonumber \\
T^{rr}&=&P(r)g^{rr}=P(r)e^{-2\Lambda} \nonumber \\
T^{\theta\theta}&=&P(r)g^{\theta\theta}=\frac{P(r)}{r^2}  \,.
\end{eqnarray}
As in Ref. \cite{Damour:2019oru} we assume that the material source is not macroscopically spin-polarized, so that we can set to zero
the direct source of the connection ${A^i}_{ j \mu}$. [As was shown in previous work \cite{Hayashi:1980ir,Nikiforova:2009qr}, 
a notable feature of the model we are considering is that the coupling to $T^{\mu\nu}$  suffices to indirectly generate a macroscopic
torsion.]

Using this material source, the field equations obtained by varying the five field variables $\{ \Lambda\,,\Phi \,, \, V\,,\, W\,,\, \pb \}$ read as follows
\begin{eqnarray}
E_{\pb}&&=
-\frac{4}{3} r \eta \lambda (\k^2 r L\,\pb + V - r F\, V - 
   L\, V - \Yb + 2 L\, \Yb \nonumber \\
   && + r L \,V\, \Yb - 
   r L \Yb^2 - r \Yb^{\prime}) \,; \label{InEqBegin}
\end{eqnarray}

\begin{eqnarray}
E_{V}&&=  -\frac{
 4 r \eta \lambda}{3 (1 + \eta)} \left[   3 r L V   - 3 r L \Yb + 3 (L-1)  \right. \nonumber \\
    && \left. + (2 \pb+ L \pb+ 
    r L \pb V- r L \pb \Yb+ 
    r \pb^{\prime})(1+\eta) \right]  \,; \nonumber \\
\end{eqnarray}

\begin{eqnarray}
E_{W}&&= -  \frac{4}{3} r^2 \eta \lambda \pb^{\prime}
 - 
 \frac{4}{3} r \eta \lambda \pb (r F + 2 L + r L V - 2 r L \Yb) \nonumber \\
 && - \frac{4 r \eta \lambda}{1+\eta} (1 + r F - L - 2 r L V + 
    r L \Yb)  \nonumber \\
    && -\frac{4}{3} r \eta \lambda \pb \,;
    \end{eqnarray}

\begin{eqnarray}
E_{\Lambda}&&=
\frac{1}{(1+\eta)L}(-2 \lambda - 4 r \lambda F + 2 \lambda L^2 + 
  r^2 L^2 P + r^2 \eta L^2 P) \nonumber \\
  && - 
 \frac{2}{3} \k^2 r^2 \eta \lambda L \pb^2 - \frac{1}{1+\eta}
 2 r \eta \lambda L V (4 + 3 r V)  \nonumber \\
 && + \frac{1}{1+\eta}
 4 r \eta \lambda L (1 + 2 r V) \Yb - \frac{1}{1+\eta}
 2 r^2 \eta \lambda L \Yb^2  \nonumber \\
&& - \frac{4}{3} r \eta \lambda L \pb (-V + 2 \Yb + r V \Yb - r \Yb^2)  \,;\label{ELambda}
\end{eqnarray}

\begin{eqnarray}
E_{\Phi}&&=
-\frac{1}{3 (1 + \eta) L}(6 \lambda - 6 \lambda L^2 + 
    3 r^2 e L^2 + 3 r^2 \eta e L^2 \nonumber \\
    && + 
    2 \k^2 r^2 \eta \lambda L^2 \pb^2 + 
    2 \k^2 r^2 \eta^2 \lambda L^2 \pb^2 + 
    24 r \eta \lambda L^2 V  \nonumber \\
    && + 
    12 r \eta \lambda L \pb V + 
    12 r \eta^2 \lambda L \pb V - 
    4 r \eta \lambda L^2 \pb V \nonumber \\
    && - 
    4 r \eta^2 \lambda L^2 \pb V + 
    18 r^2 \eta \lambda L^2 V^2 - 
    12 r \eta \lambda L \Yb \nonumber \\
    && - 
    12 r \eta \lambda L^2 \Yb - 
    4 r \eta \lambda L \pb \Yb - 
    4 r \eta^2 \lambda L \pb \Yb \nonumber \\
    && + 
    8 r \eta \lambda L^2 \pb \Yb + 
    8 r \eta^2 \lambda L^2 \pb \Yb - 
    24 r^2 \eta \lambda L^2 V \Yb \nonumber \\
    && + 
    4 r^2 \eta \lambda L^2 \pb V \Yb + 
    4 r^2 \eta^2 \lambda L^2 \pb V \Yb + 
    6 r^2 \eta \lambda L^2 \Yb^2 \nonumber \\
    && - 
    4 r^2 \eta \lambda L^2 \pb \Yb^2 - 
    4 r^2 \eta^2 \lambda L^2 \pb \Yb^2) \nonumber \\
    &&  + \frac{
 4 r \lambda L^{\prime}}{(1 + \eta) L^2} - 
 \frac{4}{3} r^2 \eta \lambda V \pb^{\prime} \nonumber \\
 && - 
 \frac{4 r^2 \eta \lambda}{3 (1 + \eta)} (3 + 
    \pb + \eta \pb) (V^{\prime} - 
    \Yb^{\prime}) \,. \label{InEqEnd}
\end{eqnarray}
These five field equations  must be supplemented by a matter equation. The latter equation follows from the radial conservation law $\nabla ^g_{\mu}T^{\mu\nu}=0$ ($\nabla ^g_{\mu}$ is the covariant derivative associated with the Levi-Civita connection), which should be written for a spherically symmetric configuration. Then the matter equation reads as follows
\be \label{Peq}
P^{\prime} + (e+P)\frac{d\Phi}{dr}=P^{\prime} + (e+P)F =0 \,.
\ee
Let us simplify the system. We already mentioned that the equation $E_{\Lambda}$ is algebraic. There is one more algebraic equation in this system, namely, the linear combination $E_{V}-E_{W}$.
These two algebraic equations can then be used to express $F$ and $L$ in terms of $V$, $\Yb$ and $\pb$. As the equation \eqref{ELambda} is quadratic in $L$, the expressions for $L$ and $F$ contain square roots, and generally there are two roots. We choose the root according to the requirement that $L$ should go to $1$ in the flat limit, see \eqref{flatlim}. So we can resolve these two algebraic equations to obtain
\bea \label{LFeq}
&L=f_{L}(V\,,\; \Yb\,,\; \pb\,,\;P\,,\; \eta\,,\; \lambda\,,\; r\,,\; \k^2\, )  \label{EqL} \\
&F=f_{F}(V\,,\; \Yb\,,\; \pb\,,\;P\,,\; \eta\,,\; \lambda\,,\; r\,,\; \k^2\, )\,.  \label{EqF}
\eea
The functions $f_L$ and $f_F$ are algebraic functions, and we have checked that  they are smooth in the limit $\k \to 0$. Differentiating  \eqref{EqL} one obtains an equation
\be
L^{\prime}=\overline{f_{L}}(V^{\prime},\, \Yb^{\prime},\,\pb^{\prime},\; V\,,\; \Yb\,,\; \pb\,,\;P^{\prime}\,,\; \eta\,,\; \lambda\,,\; r\,,\; \k^2\, )  \label{Lprime}
\ee
relating $L^{\prime}$, $V^{\prime}$, $\Yb^{\prime}$ and $\pb^{\prime}$ (each of them entering linearly), with some algebraic function of $L$, $V$, $\Yb$ and $\pb$. The function $\overline{f_{L}}$ is also smooth in the limit $\k \to 0$. 

The three remaining equations $E_{\pb}$, $E_{V}$ and $E_{\Phi}$ are differential equations containing $\pb^{\prime}$, $V^{\prime}$, $\Yb^{\prime}$ and $L^{\prime}$. Two of them, $E_{\pb}$ and $E_{V}$, can be easily solved with respect to the derivatives
of $\Yb$ and $\pb$:
\begin{eqnarray}
r\Yb^{\prime} &=&\k^2 r L \pb + (1 - r F - L) V \nonumber\\
&&+ ( 2 L-1 + r L V) \Yb - r L \Yb^2 \label{1stExEq}
\end{eqnarray}
\begin{eqnarray}
 r (1 + \eta) \pb^{\prime}&=&3 ( 1-L) - (1 + \eta) (2 + L) \pb \nonumber \\
 && - 
 r L (3 + \pb + \eta \pb) (V - \Yb)  \,. \label{2ndExEq}
\end{eqnarray}
They are obviously smooth in the limit $\k \to 0$.  

Eq. \eqref{InEqEnd} contains the derivatives $\pb^{\prime}$, $V^{\prime}$, $\Yb^{\prime}$ and $L^{\prime}$, all entering linearly.
We can substitute in Eq. \eqref{InEqEnd} the expression \eqref{Lprime} for $L^{\prime}$. This yields a differential equation which is linear in $\pb^{\prime}$, $V^{\prime}$, $\Yb^{\prime}$. This equation again has a smooth limit with $\k \to 0$. Finally, substituting 
Eqs. \eqref{1stExEq} and \eqref{2ndExEq}, we obtain an equation for $V^{\prime}$, which still has good behavior in $\k \to 0$. It reads as follows
\be
C_{V'}V^{\prime} + f_V(L, \,F,\,V\,,\Yb,\,\pb,\, P^{\prime},\,P,\,e,\,\k^2)=0 \,. \label{EqVprime}
\ee
The function $f_V$ is smooth as $\k \to 0$. The coefficient $C_{V'}$ has the structure
\be
C_{V'}=C_{V'0}+\k^2C_{V'2} \,,
\ee
and in the flat limit \eqref{flatlim} it equals
\be
C_{V'}^{\rm flatlim}=-\frac{48r^3\eta\lambda^2}{1+\eta} \;.
\ee
This means that Eq. \eqref{EqVprime} can be solved with respect to the derivative $V^{\prime} $ without introducing any dangerous denominators in the limit $\k \to 0$ (at least in the weak field domain, i.e. where the deviation from the flat limit is small).

Thus we have shown that the system of five field equations, \eqref{EqL}, \eqref{EqF}, \eqref{1stExEq}, \eqref{2ndExEq}, \eqref{EqVprime} 
(to be completed, in the matter, by Eq. \eqref{Peq}) can be replaced by a system made of the two algebraic equations \eqref{LFeq}--\eqref{EqF},
and of three first-order differential equations solved with respect to the derivatives of the three remaining variables 
$ (Y_1, Y_2,Y_3) \equiv (V\,,\; \Yb\,,\; \pb)$, say
\be
Y_a^{\prime}= f_a(Y_b, \, \eta\,,\; \lambda\,,\; r\,,\; \k^2) \,,\; \; (a=1,2,3) \,.
\ee
The right hand sides $f_a$ of these equations are analytic (actually algebraic) in  $\k^2$, and regular as $\k^2 \to 0$. Here, we have
not indicated the dependence of the $f_a$'s on the matter variables $e$ and $P$. We recall that
to complete the system, inside the star we need to augment it by the matter equation
\be
P^{\prime}=-(e+P)F \label{MatterEqV1}
\ee
(with $F$ given by \eqref{EqF}), together with some  equation of state for the star,
\be
e=e(P) \,. \label{Eqrho}
\ee
In view of general theorems on ordinary differential equations smoothly dependending on parameters, this shows that, given 
some initial conditions for $Y_a(r_0, \k^2)$ (smooth in $\k^2$) at some radius $r=r_0$, there will exist, in some neighbourhood
of $r_0$, a solution $Y_a(r, \k^2)$, smooth in $\k^2$. By adapting the discussion in Section IX of \cite{Damour:2019oru},
one can actually take $r_0=0$, and (given some equation of state for the matter ) so define a smooth-in-$\k^2$ solution  $Y_a(r, \k^2)$,
depending on the choice of a unique datum, say the value $v_1$ of $V^{\prime}$ at the origin $r=0$. As $\k$ has the
dimension of an inverse length (that we can take to be much larger than the radius of the star), we physically expect that the
so-constructed solution will be well-defined in the whole domain $\k r \lesssim 1$. However, this does not prove the global existence
of a smooth-in-$\k^2$, physically acceptable solution because we would like to impose the boundary condition that this solution
tends to the flat limit at infinite distances $r \to \infty$. The existence of asymptotically flat solutions was
numerically proven in Ref. \cite{Damour:2019oru} (by appropriately shooting the unique parameter $v_1$) for the original
system of field equations (containing $1/\k^2$ factors). We leave to future work an investigation of numerical solutions
to our new, smooth-in-$\k^2$, system. In the next section, we will appeal to perturbative theory to study the existence of
globally regular, smooth-in-$\k^2$, solutions.

\section{Perturbation theory} \label{PT}

The system \eqref{EqL}, \eqref{EqF}, \eqref{1stExEq}, \eqref{2ndExEq}, \eqref{EqVprime} is rather complex. Let us, however, show how one can solve it by successive approximations around the weak field limit \eqref{flatlim}. Let us look for solutions of our system in the form of a nonlinearity expansion
\begin{eqnarray}
V&=& \sum_{n=1}^{\infty}V_n \,, \quad \Yb= \sum_{n=1}^{\infty}\Yb_n \,, \quad F= \sum_{n=1}^{\infty}F_n \, ,\quad \nonumber \\
\pb&=& \sum_{n=1}^{\infty}\pb_n \,, \quad \overline{L}\equiv L-1=  \sum_{n=1}^{\infty}L_n \,, \label{decompose}
\end{eqnarray}
where $V_1$, $\Yb_1$, $F_1$, $\overline{L}_1$, $\pb_1$ will be the linearized solution generated by the material source. 

It is convenient to introduce the variables $V^{m0}$ and $V^{mk}$ defined as \cite{Damour:2019oru}
\bea 
V^{m0} &\equiv& -3 V+ 2 \Yb \,,\nonumber \\
V^{mk} &\equiv&  2 V - \Yb \,,
\eea
or, in reverse
\bea \label{eqY1V1}
\Yb &=& 2 V^{m0}+ 3 V^{mk} \,,\nonumber \\
V &=&  V^{m0}+ 2 V^{mk} \,.
\eea
This decomposition will be used in each order of nonlinearity, for example, $\Yb_n= 2 V^{m0}_n+ 3 V^{mk}_n$.

Let us take the system
\bea
&&E_{\pb}(\Yb^{\prime},\,\Lb,\,F, \,V,\,\Yb,\,\pb) \\
&&E_{V}(\pb^{\prime},\,\Lb, \,V,\,\Yb,\,\pb) \\
&&E_{\Phi}(\pb^{\prime},\,\Yb^{\prime},\,V^{\prime},\,\Lb^{\prime},\,\Lb, \,V\,,\Yb,\,\pb,\, e) \\
&&E_{\Lambda}(\Lb,\,F, \,V,\,\Yb,\,\pb,\,P) \\
&&[E_{V}-E_{W}](\Lb,\,F, \,V,\,\Yb,\,\pb) \;.
\eea
Let us make in these equations the change of variables \eqref{eqY1V1}. After that, we separate in these equations the terms which are linear in $V^{m0}$, $V^{mk}$, $F$, $\Lb$ and $\pb$ from the nonlinear terms. Then, after some algebraic transformations of the obtained system, we come to a system of the form
\begin{eqnarray}
&& {V^{m0}}^{\prime} +\frac{2}{r}V^{m0} \nonumber \\
&&= \frac{e-3P-rP^{\prime}}{4\lambda} + {N^{m0}}  \,, \label{Em0} \\
&& {V^{mk}}^{\prime} - \frac{\k^2}{3}\pb-\frac{1}{r}V^{mk}-\frac{2}{r}V^{m0}\nonumber \\
&&=-\frac{e-3P-rP^{\prime}}{6\lambda} + {N^{mk}} \,, \label{Emk} \\
&& {\pb}^{\prime}+\frac{3}{r}\pb-3V^{mk}= -\frac{3}{4\lambda}rP + {N^{\pb}}  \,, \label{Ep} \\
&&F= {V^{m0}} - 2 \eta {V^{mk}}+\frac{r(1 +  \eta) P}{2 \lambda} + {N^F}  \,, \\
&&\Lb= r{V^{m0}} - r \eta {V^{mk}}+\frac{r^2(1 +  \eta) P}{4 \lambda} + {N^L}  \,. \label{ELb}
\end{eqnarray}
Here $N^{\pb}$, $N^{F}$, $N^{L}$, $N^{m0}$ and $N^{mk}$ are nonlinear functions of the variables $V^{m0}$, $\pb$, $V^{mk}$, $\Lb$, $F$, and their derivatives. [The $N^a$'s are at least quadratic in $V^{m0}$, $V^{mk}$, $\pb$,  $\Lb$, $F$ (or, equivalently $V$, $\Yb$, $\pb$,  $\Lb$, $F$) and their derivatives.]

Taking the linear combination $3 \times \eqref{Emk}+r\times \left[\eqref{Ep}/r\right]^{\prime}$ we obtain 
the following third-order differential system for the two variables (${V^{m0}}, \pb$) 
\begin{eqnarray}
&&  {V^{m0}}^{\prime} +\frac{2}{r}{V^{m0}}={S^{m0}} \,, \label{Sysn1} \\
&& {\pb}^{\prime\prime}+\frac{2}{r}{\pb}^{\prime} - \left(\frac{6}{r^2}+\k^2\right)\pb ={S^{\pb}} \,,  \label{Sysn2} 
\eea
where the corresponding source terms $(S^{m0}, {S^{\pb}})$ read
\begin{eqnarray}
&& {S^{m0}} \equiv \frac{e-3P-rP^{\prime}}{4\lambda} + {N^{m0}} \,, \label{Sm0} \\
 &&{S^{\pb}} \equiv  \frac{6}{r}V^{m0} - \frac{2e-6P+rP^{\prime}}{4\lambda} \\
 &&+r\left(\frac{{N^{\pb}}}{r}\right)^{\prime} +\,3{N^{mk}} \nonumber \\
 &&= \frac{6}{r}V^{m0} -\frac{2e-6P+rP^{\prime}}{4\lambda} + \widehat{N}^{\pb}  \,,
 \end{eqnarray}
 and
 $$
 \widehat{N}^{\pb} \equiv r\left(\frac{{N^{\pb}}}{r}\right)^{\prime} +\,3{N^{mk}} \,.
 $$
Note that one needs to first obtain $V^{m0}$ by solving Eq. \eqref{Sysn1} so as to insert it in the source term ${S^{\pb}}$ for $\pb$.

After having determined ${V^{m0}}, \pb$, one then successively computes $ V^{mk}$, $F$ and $\Lb$ by the equations
\bea
&& V^{mk}=\frac{1}{3}\left( {\pb}^{\prime} +\frac{3}{r}\pb  \right)+\frac{rP}{4\lambda}-\frac{1}{3}{N^{\pb}} \label{Sysn3} \,, \\
&&F= {V^{m0}} - 2 \eta {V^{mk}} +\frac{r(1 +  \eta) P}{2 \lambda} + {N^F} \,, \label{EqFn} \\
&&\Lb= r{V^{m0}} - r \eta {V^{mk}} +\frac{r^2(1 +  \eta) P}{4 \lambda} + {N^L} \,. \label{EqLn}
\end{eqnarray}


The equations \eqref{Sysn1}--\eqref{EqLn}, together with the material equation \eqref{MatterEqV1}--\eqref{Eqrho}, can be solved by successive iterations. 

First, at the linear order, the system reads
\begin{eqnarray}
&& {V^{m0}_1}^{\prime} +\frac{2}{r}V^{m0}_1 =S^{m0}_1\equiv \frac{e-3P-rP^{\prime}}{4\lambda} \,, \label{Sys11} \\
&& {\pb_1}^{\prime\prime}+\frac{2}{r}{\pb_1}^{\prime} - \left(\frac{6}{r^2}+\k^2\right)\pb_1 \nonumber \\
&&=S^{\pb}_1 \equiv \frac{6}{r}V^{m0}_1-\frac{2e-6P+rP^{\prime}}{4\lambda} \,, \label{Sys1pb}\\
&& V^{mk}_1=\frac{1}{3}\left( {\pb_1}^{\prime} +\frac{3}{r}\pb_1 +\frac{3}{4\lambda}rP \right) \,, \nonumber \\
&&F_1= V^{m0}_1 - 2 \eta V^{mk}_1+\frac{r(1 +  \eta) P}{2 \lambda} \,, \label{Sys12p} \\
&&\Lb_1= rV^{m0}_1 - r \eta V^{mk}_1+\frac{r^2(1 +  \eta) P}{4 \lambda}  \;. \label{Sys13}
\end{eqnarray}
Here $S^{\pb}_1$ and $S^{m0}_1$ in the linear order contain only the material source ($V_1^{m0}$ in $S^{\pb}_1$
being itself determined in terms of the material source by solving the first equation). 

As a next step, one can consider the second order in nonlinearity, i.e. including the quadratic terms. The source terms in \eqref{Sysn1}--\eqref{EqLn} will contain no material sources\footnote{For conceptual simplicity, we assume here that the material contributions which
depend on the pressure $P$ are incorporated in the first-order solution, though, as will be discussed, $P$ is a second-order quantity.} but terms quadratic in field variables. In agreement with the idea of iteration, we substitute in these quadratic source terms the first-order solution. We will denote the quantities generated by quadratic source terms as $V^{m0}_2$, $\pb_2$ etc, and they will satisfy the following equations
\begin{eqnarray*}
&& V^{m0\,\prime}_2 +\frac{2}{r}V^{m0}_2=S^{m0}_2 \,, \nonumber \\
&& {\pb}_2^{\prime\prime}+\frac{2}{r}{\pb}_2^{\prime} - \left(\frac{6}{r^2}+\k^2\right)\pb_2 =S^{\pb}_2 \,,\nonumber \\
&& V^{mk}_2=\frac{1}{3}\left( {\pb}_2^{\prime} +\frac{3}{r}\pb_2  \right)-\frac{1}{3}N^{\pb}_2  \,, \\
&&F_2= V^{m0}_2 - 2 \eta V^{mk}_2 + N^F_2  \,, \\
&&\Lb_2= rV^{m0}_2 - r \eta V^{mk}_2  + N^L_2 \,,
\end{eqnarray*}
where $S^{m0}_2$  is quadratic in  $V^{m0}_1$, ${\pb}_1$, $V^{mk}_1$  etc.; $S^{\pb}_2 \equiv  \frac{6}{r}V^{m0}_2 + \widehat{N}^{\pb}_2$, while $N^{\pb}_2$, $\widehat{N}^{\pb}_2$, $N^F_2$ and $N^L_2$ are quadratic functions in $V^{m0}_1$, $V^{mk}_1$, $\pb_1$, $F_1$ and $\Lb_1$. This iteration process  can be continued to  higher orders. 

Generally, given any sources $S^{\pb}$ and $S^{m0}$ (regular at the origin, and decaying sufficiently fast at radial infinity),
there is a {\it unique} solution of the equations \eqref{Sysn1}--\eqref{Sysn2} which is correspondingly regular at the origin
and decaying at infinity (see the analog discussion in \cite{Damour:2019oru}). It is obtained as follows. First, the origin-regular
 solution of Eq. \eqref{Sysn1} is obtained by the simple integral
\be
{V^{m0}}(r)=\frac{1}{r^2}\int_{0}^{r}\hat{r}^2{S^{m0}}(\hat{r})d\hat{r}  \,.
\ee
Then we substitute this result in the source term of Eq. \eqref{Sysn2}. 
The solution of the latter equation is then obtained by a Green's function technique:
\be
\pb(r)=\int_{0}^{\infty}\hat{r}^2G_{\kappa}(r,\hat{r}){S^{\pb}}(\hat{r})d\hat{r} \,, \label{pbgenGreen}
\ee
where the Green's function $G_{\kappa}(r,\hat{r})$ is constructed using two homogeneous solutions, $X_{>}(r)$ and $X_{<}(r)$, 
of the $\pb$ equation in the following way
\bea
G_{\kappa}(r,\hat{r}) &&\equiv \frac1{\mathcal{W}}\left[ X_{>}(r)X_{<}(\hat{r})\theta(r-\hat{r}) \right. \nonumber \\ 
&& \left. +  X_{<}(r)X_{>}(\hat{r})\theta(\hat{r}-r) \right] \,. \label{GreenFunc}
\eea
Here $\mathcal{W}$ is the conserved Wronskian of these two homogeneous solutions
\be
\mathcal{W}\equiv r^2 \left( X_{>}^{\prime}(r) X_{<}(r)-  X_{>}(r) X_{<}^{\prime}(r) \right) \label{Wronsk} \,.
\ee
The choice of homogeneous solutions is uniquely fixed by the boundary conditions at $r=0$ and $r \to \infty$ that we want the inhomogeneous solution \eqref{pbgenGreen} to satisfy. The function $X_<(r)$ need to be chosen so as to be regular at the origin,
 while $X_>(r)$ should be chosen so as to decay as $r \to \infty$.
 In addition, it is convenient to normalize our homogeneous solution such that their Wronskian is equal to 1,
$ \mathcal{W} =1$.

The homogeneous solutions of the equation \eqref{Sysn2} are given by linear combinations of the spherical Bessel functions, 
$j_{-3}(i k r)$ and $y_{-3}(i k r)$  (where $i^2=-1$).  
The unique combinations satisfying our required boundary conditions are the following.

 The outer homogeneous $X_{>}(r)$ reads
\begin{eqnarray}
X_{>}(r) &=& \k^3 \left[i \, j_{-3}(i\k r) - y_{-3}(i \k r)\right]\nonumber \\
&=&-e^{-\k r}\frac{3+3\k r+\k^2 r^2}{r^3} \,.  \label{Xout}
\end{eqnarray}
It decays exponentially in the domain $\k r \gg 1$, as one can see from \eqref{Xout}.
Note that $X_>(r)$ has a singular powerlaw behavior in the domain $\k r \ll 1$:
\be
 X_{>}(r)= -\frac{3}{r^3}+O\left(\frac{\k^2}{r}\right)  \quad (\k r \ll 1) \,. \label{X>Near0}
\ee
The inner homogeneous $X_{<}(r)$ is
\begin{eqnarray}
X_{<}(r) &=& - \frac{1}{\k^2}y_{-3}(i \k r) \nonumber \\
&=& -\frac{3 \cosh(\k r)}{\k^4 r^2}+\sinh(\k r)\frac{3+\k^2 r^2}{\k^5 r^3} \,.
\end{eqnarray}
It is regular at the origin and admits the following form in the domain $\k r \ll 1$
\be
X_{<}(r) = \frac{r^2}{15}+O(\k^2r^4)  \quad (\k r\ll 1)\,. \label{X<Origin}
\ee

In the domain $\k r \gg 1$ it grows exponentially
\be
X_{<}(r) \sim \frac{e^{\k r}}{2\k^3 r}\l 1-\frac{3}{\k r}+\frac{3}{(\k r)^2} \r \quad   (\k r\gg 1)\,. \ee

Note that both $X_>(r)$ and $X_<(r)$ have finite limits when $\k \to 0$ at fixed $r$, which means that they have 
a finite massless limit in the whole domain $\k r \ll 1$ for small $\k$.

To proceed with the construction of solutions, let us consider in detail the first order of perturbation theory.

\section{Linear order}\label{Linear}
At the linear order we have the system \eqref{Sys11} - \eqref{Sys13}. After solving that system, we will  compute $V_1$ and $\Yb_1$
by means of
\bea
\Yb_1 &=& 2 V^{m0}_1+ 3 V^{mk}_1 \,, \\
V_1 &=&  V^{m0}_1+ 2 V^{mk}_1 \,.
\eea

Let us construct the complete solution in the linear order. 
The solution for $V^{m0}$ we know from our previous paper \cite{Damour:2019oru}. It reads
\be
V^{m0}_1=\frac{m_1(r)}{r^2} + V^{m0}_P\,, \label{SolVm01}
\ee
where
\be
m_1(r) \equiv 4\pi G_0 \int_{0}^{r} \hr^2 e(\hr)d\hr = \frac{1}{4\lambda}\int_{0}^{r} \hr^2 e(\hr)d\hr
\ee
and $V^{m0}_P$ is a pressure contribution which will be discussed later in this section, and which we do not write
explicitly at this stage of the calculation (because it is actually of the second order). 

The solution for $\pb_1$ is constructed using the Green's function discussed above. It reads, when
considering the  interior solution $r < R_S$ (where $R_S$ denotes the radius of the star)
\begin{eqnarray}
\pb_1^{\;(r<R_S)}&=&-2X_{>}(r)\int_{0}^{r}X_{<}(\hr)dm_1(\hr) \nonumber \\
&& - 2X_{<}(r)\int_{r}^{R_S}X_{>}(\hr)dm_1(\hr) \nonumber \\
&&+ 6X_{>}(r)\int_{0}^{r}X_{<}(\hr)\frac{m_1(\hr)}{\hr}d\hr  \nonumber \\
&& + 6X_{<}(r)\int_{r}^{R_S}X_{>}(\hr)\frac{m_1(\hr)}{\hr}d\hr \nonumber \\
&& - 6 m_1 e^{-\k R_S}\frac{1+\k R_S}{R_S^3} X_{<}(r) + \pb_P^{(r<R_S)}\,. \nonumber \\ \label{pb<R_S}
\end{eqnarray}
Here, and in the following, we use the notation $m_1$ to denote $m_1(R_S)$, {\it i.e.}, $G_0 M$, where $$ M \equiv 4\pi \int_0^{R_S} \hr^2 e(\hr) d\hr $$ is the total mass-energy of the star.  

In the exterior of the star, the solution for $\pb_1$ reads
\begin{eqnarray}
\pb_1^{\;(r>R_S)}&=&-2X_{>}(r)\int_{0}^{R_S}X_<(\hr)dm_1(\hr) \nonumber \\
&&+ 6X_>(r) \int_{0}^{R_S}X_<(\hr)\frac{m_1(\hr)}{\hr}d\hr \nonumber \\
&& +6m_1X_>(r)\left[ \frac{\k\hr \cosh(\k \hr) - \sinh(\k \hr)}{\k^5 \hr^3} \right]\bigg|_{R_S}^r \nonumber \\
&& - 6m_1 e^{-\k r}\frac{1+\k r}{r^3}X_<(r)+ \pb_P^{(r>R_S)} \,.\label{pb>R_S}
\end{eqnarray}
Here $\pb_P^{(r<R_S)}$, $\pb_P^{(r>R_S)}$  are pressure contributions which will be discussed below.  The last expression 
(without the pressure term) can be rewritten in a more convenient manner, namely
\be
\pb_1^{\;(r>R_S)}=C_0(\k,R_S)X_>(r) - \frac{6m_1}{\k^2 r^3} \label{pbFirst}
\ee
with
\begin{eqnarray}
&&C_0(\k, R_S) \equiv -2\int_{0}^{R_S}X_<(\hr)dm_1(\hr) \nonumber \\
&& + 6\int_{0}^{R_S}X_<(\hr)\frac{m_1(\hr)}{\hr}d\hr \nonumber \\
&&- 6m_1\frac{\k R_S \cosh(\k R_S) - \sinh(\k R_S)}{\k^5 R_S^3} \nonumber \\
&& = -2\int_{0}^{R_S}X_<(\hr)dm_1(\hr) + 6\int_{0}^{R_S}X_<(\hr)\frac{m_1(\hr)}{\hr}d\hr   \nonumber \\
&& - \frac{2m_1}{\k^2}{\cal F}(\k R_S) \,, \label{Cbar}
\end{eqnarray}
where the form factor ${\cal F}(z)$ was introduced in our previous paper \cite{Damour:2019oru} and reads
\be
{\cal F}(z) \equiv 3\left\{  z \,\cosh{z} - \sinh{z}  \right\}/z^3 \,. \label{defF1}
\ee
To clarify the structure of $C_0$, let us introduce the function $C(\k, R_S)$ as follows
\be
C_0 \equiv -\frac{2m_1}{\k^2} - 3C(\k, R_S) \,.
\ee
Then $C(\k, R_S)$ reads, according to Eqs. \eqref{Cbar} and \eqref{defF1},
\bea
&&-3C = -2\int_{0}^{R_S}X_<(\hr)dm_1(\hr) \nonumber \\
&& + 6\int_{0}^{R_S}X_<(\hr)\frac{m_1(\hr)}{\hr}d\hr \nonumber \\
&& -2m_1 R_S^2 \overline{{\cal F}}(\k R_S) \,, \label{ConstC}
\eea
where we introduced a new form factor:
\be
\overline{{\cal F}}(z_S) \equiv \frac{3(z_S\,\cosh{z_S}-\sinh{z_S})-z_S^3}{z_S^5} \,.
\ee
Let us consider the case where $\k \ll R_S^{-1}$ (as appropriate when studying the $\k \to 0$ limit). 
Since the integration variable $\hr$ takes its values in the interval $0 \leq \hr \leq R_S$, we can consider the series expansion of \eqref{ConstC} in powers of $\k$ (using $\k \ll m^{-1}$ which follows from $\k \ll R_S^{-1}$ as we are considering a 
weakly self-gravitating star). The important remark is that, in spite of the apparent $1/z_S^5$ singular denominator in $z_S=\k R_S \to 0$, 
the form factor $\overline{{\cal F}}(z_S) $ is regular as $z_S = \k R_S \to 0$, namely
\be
\overline{{\cal F}}(z_S) = \frac{1}{10}+O(z_S^2) \,. 
\ee
Thus, $C(\k, R_S)$ is finite in the limit $\k R_S \ll 1$. 

Let us now show that, in turn, this finiteness of $C(\k, R_S)$ in the $\k \to 0$ limit, implies that 
 $\pb_1^{\;(r>R_S)}$ has a finite limit as $\k \to 0$, because of a cancellation between the $O(\k^{-2})$ terms
 present both in $C_0$ and in the $- \frac{6m_1}{\k^2 r^3}$ contribution to  $\pb_1^{\;(r>R_S)}$, in Eq. \eqref{pbFirst}.
Indeed, Eq. \eqref{pbFirst} can be rewritten as follows
\bea
\pb_1^{\;(r>R_S)}&&= -3CX_>(r) +\pb^{\rm add}
  \nonumber \\
&&= 3C\frac{3+3\k r+\k^2 r^2}{r^3}e^{-\k r} +\pb^{\rm add} \,. \label{pblin}
\eea
Here
\bea
\pb^{\rm add} &&\equiv \frac{2m_1}{\k^2}\frac{3+3\k r+\k^2 r^2}{r^3}e^{-\k r} - \frac{6m_1}{\k^2 r^3} \label{pbSecond} \nonumber \\
&& = \frac{2m_1}{r}{\cal F}^{\pb}(\k r)\,, \label{pbadd}
\eea
where we defined
\bea
&&{\cal F}^{\pb}(z)\equiv \frac{3+3z+z^2}{z^2}e^{-z}-\frac{3}{z^2}\,, \\
&& {\cal F}^{\pb}(z) = -\frac{1}{2}+O(z^2) \quad(z\ll 1)\,.
\eea
Therefore, though  the expression \eqref{pbSecond} for $\pb^{\rm add}$  contains some explicit $\k^{-2}$ factors, $\pb^{\rm add}$
 is actually regular in the  $\k \to 0$ limit, so that we have
$$ \pb^{\rm add} = -\frac{m_1}{r} + O(m_1 r \k^2) \,.$$

The function ${\cal F}^{\pb}(z)$ (where $z = \k r$) is finite everywhere including $z \to 0$ (corresponding to $\k\to0$), and decays as   $z \to \infty$ (i.e. $r\to\infty$). In the domain $r \gg\k^{-1}$ one has
\be
\pb^{\rm add}(r) \approx - \frac{6m_1}{r}\frac{1}{\k^2r^2} \ll \frac{m_1}{r} \,.
\ee
Therefore, though there remains an explicit factor $\k^{-2}$ in the large-$r$ expression of  $\pb^{\rm add}(r) $,
this factor does not imply any growing behavior in the $\k \to 0$ limit, because this factor only appears 
in the domain  $r \gg\k^{-1}$, so that one still gets the $\k$-independent bound 
$\pb^{\rm add}(r)  \ll \frac{m}{r} $ for $\pb^{\rm add}(r)$. In that sense the  $\k \to 0$ limits
of both $\pb^{\rm add}(r)$ and $\pb(r)$ are finite.


Finally, from \eqref{Sys13} one can compute $V^{mk}_1$
\begin{eqnarray}
V^{mk}_1&=&\frac{1}{3}\left( {\pb_1}^{\prime}+\frac{3}{r}\pb_1 \right)\nonumber \\
&=&\frac{C_0(\k, R_S)\k^2}{3}e^{-\k r}\frac{1+\k r}{r^2}\nonumber \\
&=&-\left( \frac{2m_1}{3}+C(\k, R_S) \k^2 \right)e^{-\k r}\frac{1+\k r}{r^2} \,. \nonumber \\ \label{Vmklin}
\end{eqnarray}

We recall that when writing the structure \eqref{Vmklin} we omitted  pressure contributions $V^{m0}_P$ and $\pb_P$ in \eqref{SolVm01} and \eqref{pb<R_S}, \eqref{pb>R_S}. Let us now consider the corrections to the exterior solution coming from taking into account these pressure terms. 

First, let us consider the equation \eqref{Sys11}. There we can incorporate pressure terms defining a corrected value of $e(r)$, namely
\bea
\hat{e}(r) &&\equiv e(r) - 3P(r)-rP^{\prime}(r) \, ,\\
e(r) &&= \hat{e}(r) + 3P(r)+rP^{\prime}(r) \,.
\eea
Then the exterior solution with pressure terms taken into account will have the same form as \eqref{SolVm01} but with a corrected mass
\be
V^{m0}_1(r) = \frac{m_1(r)}{r^2}+V^{m0}_P(r)=\frac{m_1(r)+\delta_Pm(r)}{r^2}
\ee
where 
\be
\delta_Pm(r) \equiv  \frac{1}{4\lambda}\int_{0}^{r} \hr^2 [-3P(\hr) - \hr P^{\prime}(\hr)]d\hr  \,.
\ee
Let us look at the equation for $\pb_1$ \eqref{Sys1pb}. It rewrites in the following form
\bea
&& {\pb_1}^{\prime\prime}+\frac{2}{r}{\pb_1}^{\prime} - \left(\frac{6}{r^2}+\k^2\right) = \frac{6}{r}V^{m0}_1(\hat{e}) - \frac{2\hat{e}}{4\lambda} - \frac{3rP^{\prime}}{4\lambda} \,. \nonumber\\
\eea
This is the same equation as the Eq.~\eqref{Sys1pb} with $P\equiv0$, {\it except for the last term} 
$$ - \frac{3rP^{\prime}}{4\lambda}
 $$
which gives an additional source term. Let us consider the equation with only this additional source term
\be
{\pb_1}^{\prime\prime}+\frac{2}{r}{\pb_1}^{\prime} - \left(\frac{6}{r^2}+\k^2\right) =- \frac{3rP^{\prime}}{4\lambda} \,. \label{pbPeq}
\ee 
This source term is localized, thus an exterior solution of \eqref{pbPeq} will be given by a homogeneous solution. Indeed, using the Green's function, we can write
\bea
&&\pb^{(r>R_S)}_{P}(r)=\int_{0}^{\infty}\hat{r}^2G_{\kappa}(r,\hat{r})\l- \frac{3\hr P^{\prime}(\hr)}{4\lambda}\r d\hat{r}  \nonumber \\
&&=\int_{0}^{\infty}\hat{r}^2\left[ X_{>}(r)X_{<}(\hat{r})\theta(r-\hat{r})  \right. \nonumber \\
&& \left.+ X_{<}(r)X_{>}(\hat{r})\theta(\hat{r}-r) \right] \l- \frac{3\hr P^{\prime}(\hr)}{4\lambda}\r d\hat{r}  \nonumber \\
&&= -\frac{3}{4\lambda}X_>(r)\int^{R_S}_{0}\hr^3 P^{\prime}(\hr)X_<(\hr)d \hr  \label{pbCorrP} \\
&&=\const X_>(r) \,. \nonumber 
\eea
 
Thus the correction to $\pb_1$ coming from the pressure terms will just modify the constant $C_0$ in \eqref{pbFirst}. More precisely, since \eqref{pbCorrP} is finite in the limit $\k\to0$ with fixed $r$, only $C$ \eqref{ConstC} will be modified. 

As a conclusion, taking $P$-terms into account in \eqref{Sys11}--\eqref{Sys13} will modify the exterior solution only by modifying  the 
values of the parameters $m_1$ and $C$. In addition, these  $P$-modifications are actually of the 
next order in the perturbation (i.e. in nonlinearity) expansion (in comparison with the first order in the energy density $e$).

Indeed, let us consider $e(r)$ as a primary source of all the field variables. Then defining a bookkeeping parameter $\varepsilon$ counting the order in the weak-field expantion, we can write $e=\varepsilon e_1 $, and then it defines the order of expansion for all the variables: $F=\varepsilon F_1+O(\varepsilon^2)$ etc., including $P(r)$. The latter is related to $e(r)$ by the matter equation \eqref{MatterEqV1}. This gives 
\begin{eqnarray}
P^{\prime}&=&-(e+P)F \nonumber \\
&=& -(\varepsilon e_1 + P)(\varepsilon F_1 + O(\varepsilon^2)) \,.
\end{eqnarray}
Taking into account the fact that $P(r)$ vanishes at the surface of the star we conclude that the pressure $P$ is second-order in $\varepsilon$: $P=P_2 \varepsilon^2 +O(\varepsilon^3)$. This means that the contributions linked $P$-source term in the linearized
field equations actually contribute to the second order of our perturbation theory. More precisely, the $P$-corrections in the exterior solution
only bring  second-order corrections to the first-order parameters $m_1$ and $C$. For this reason, in the discussion of the next sections, we will set $P$ to zero for simplicity. 

Let us also briefly discuss the $P$-corrections to the interior solution. The interior corrections $V^{m0}_P$ and $\pb_P$ are as follows
\bea
&&  V^{m0}_P = \frac{\delta_Pm(r)}{r^2} \,, \nonumber \\
&&   \nonumber \\
&&  \pb_P^{(r<R_S)} = -\frac{3}{4\lambda}X_>(r)\int_0^r \hr^3 P^{\prime}(\hr)X_<(\hr)d \hr \nonumber \\
&& - \frac{3}{4\lambda}X_<(r)\int_r^{R_S} \hr^3 P^{\prime}(\hr)X_>(\hr)d \hr \nonumber \\
&&+ \pb_1^{\;(r<R_S)}(\delta_P m(r)) \,, \nonumber 
\eea
where $\pb_1^{\;(r<R_S)}(\delta_P m(r))$ is the expression \eqref{pb<R_S} without $\pb_P^{(r<R_S)}$ and with 
$m_1(r)$ replaced by $\delta_Pm(r)$.

It can be easily seen that the solution found in the previous paper \cite{Damour:2019oru} has indeed the same structure as \eqref{Vmklin} (i.e. the constant in the solution for $V^{mk}_1$ has form $-\frac{2m}{3} - \k^2 C (\k, R_S)$, see there the Eqs. (7.28) -- (7.32)). Finally, let us mention that, for the constant-density case, $e(r)=\const$, the solution \eqref{Cbar}, \eqref{Vmklin} coincides
with the linear constant-density solution  found in \cite{Damour:2019oru}.

\section{Structure of the all-order perturbative exterior solution in the massless case $\k=0$}

In this section we will prove that, in the massless limit $\k=0$, one can perturbatively construct (to all nonlinearity orders) a regular complete 
 solution \eqref{decompose} which decays (in a power-law manner) at $r\to\infty$. We will prove that this solution is well-defined everywhere. The fact that one can construct such a solution by using the  $\k\to 0$ limit of the unique Green's function $G_\k(r,r')$ incorporating the physically required
 boundary conditions (both at $r=0$ and at $r=\infty$) suggests that the all-order solution \eqref{decompose} in
 the  $\k \neq 0$ case can also be constructed and that it satisfies the required boundary conditions,
 and has also a smooth limit when $\k\to0$.

The system \eqref{Sysn1}--\eqref{EqLn}, prepared for nonlinear iteration of second order and higher, reads for $\k=0$ as follows
\begin{eqnarray}
&& {V^{m0}}^{\prime} +\frac{2}{r}{V^{m0}}=N^{m0}_{\k=0} \,, \label{Sysnn1} \\
&& {\pb}^{\prime\prime}+\frac{2}{r}{\pb}^{\prime} - \frac{6}{r^2}\pb =\frac{6}{r}V^{m0} +{\widehat{N}^{\pb}}_{\k=0} \,,  \label{Sysnn2} \\
&& V^{mk}=\frac{1}{3}\left( {\pb}^{\prime} +\frac{3}{r}\pb  \right)-\frac{1}{3}N^{\pb}_{\k=0} \,, \label{Sysnn3} \\
&&F= {V^{m0}} - 2 \eta {V^{mk}}  + N^F_{\k=0} \,, \label{EqFnn} \\
&&\Lb= r{V^{m0}} - r \eta {V^{mk}} + N^L_{\k=0} \,, \label{EqLnn}
\end{eqnarray}
where $N^{m0}_{\k=0}$, $\widehat{N}^{\pb}_{\k=0}$ etc are the functions ${N^{m0}}$, ${\widehat{N}^{\pb}}$ etc from \eqref{Sm0}--\eqref{EqLn} where one takes $\k=0$.

The homogeneous solutions for the case of $\k=0$ are (see \eqref{X>Near0} and \eqref{X<Origin})
\be
X_{>}(r) = -\frac{3}{r^3} 
\ee
and 
\be
 X_{<}(r)= \frac{r^2}{15} \,.
\ee



At this stage it is convenient to work with dimensionless variables and with equations in dimensionless form. We recall that, in our treatment, the variables $V$, $\Yb$, $V^{m0}$, $V^{mk}$ and $F$ have dimension $r^{-1}$, while the variables $\Lb$ and $\pb$ are dimensionless.  Let us define corresponding dimensionless variables
\bea
&& Z_{(m0)} \equiv V^{m0} r \,, \;  Z_{(mk)} \equiv V^{mk} r \,, \;  Z_{(\Yb)} \equiv \Yb r \,,\; Z_{(V)} \equiv  V r \,, \; \nonumber \\
&&Z_{(F)} \equiv  r F  \,,\;    Z_{(\Lb)} \equiv \Lb  \,,\; Z_{(\pb)}  \equiv   \pb   \;. \label{Zdef}
\eea
Using the linear solution in the form \eqref{SolVm01}, \eqref{pblin}--\eqref{pbadd} and \eqref{Vmklin}, and the relations \eqref{Sys12p}, \eqref{Sys13} (putting in the latter ones $P=0$) we can compute the linear solution in the limit $\k=0$. The exterior solution, presented in dimensionless form, looks as follows
\begin{align}
& Z_{(m0)1}=\frac{m_1}{r}\,,\;\;  Z_{(mk)1}=-\frac{2m_1}{3r} \,, & \nonumber \\
&Z_{(V)1}=-\frac{m_1}{3r}\,,\;\; Z_{(\Yb)1}=0\,, \; &Z_{(\Lb)1}=\frac{m_1(3+2\eta)}{3r} \,, \nonumber \\
&Z_{(F)1}= \frac{m_1(3+4\eta)}{3r}  \,, \; &Z_{(\pb)1}=\frac{9  C }{r^3} - \frac{m_1}{r} \,. \label{K0Order1}
\end{align}
As one can see, the first-order solution is polynomial in $1/r$ and generally can be presented in the form of
\be
Z_{(i)1}=a_{(i)1}\frac{m_1}{r} + a_{(i)2}\frac{ C }{r^3} \,, \; (i=m0,mk,V,\Yb,\Lb,F,\pi) \,,
\ee
where $a_{(i)1}$, $a_{(i)2}$ are some numerical coefficients depending on $\eta$. Then it means that for derivatives we have the same structure, namely
\be
r Z_{(i)1}^{\prime}=b_{(i)1}\frac{m_1}{r} + b_{(i)2}\frac{ C }{r^3} \,,
\ee
with other coefficients $b_{(i)1}$, $b_{(i)2}$ (actually, $b_{(i)1}=-a_{(i)1}$ and $b_{(i)2}=-3a_{(i)2}$). 
To avoid explicitly introducing coefficients of the type of  $a_{(i)1}$, $a_{(i)2}$ and $b_{(i)1}$, $b_{(i)2}$,
it will be henceforth convenient to use the special notation ``\&'' introduced by Penrose (as cited
in Ref. \cite{Thorne:1984mz} page 1822). The notation ``\&'' means ``and a term of the form". Using this notation, we can rewrite the last two statements in the simplified form
\bea
Z_{(i)1}&\sim& \frac{m_1}{r}\&\frac{ C }{r^3} \,,\nonumber \\
r Z_{(i)1}^{\prime}&\sim&\frac{m_1}{r}\&\frac{ C }{r^3} 
\,. \label{FeatureZ1}
\eea

Now let us consider the initial system of equations \eqref{InEqBegin}-\eqref{InEqEnd} {\it setting $\k=0$ everywhere}. It is easy to check that each equation in the system \eqref{InEqBegin}-\eqref{InEqEnd}, written in the dimensionless variables \eqref{Zdef}, takes the following dimensionless form
\begin{align}
&&r Z^{\prime}_{(i)}+ c_{ij}Z_{(j)} = a_{ijk}Z_{(j)} Z_{(k)} + b_{ijkl}Z_{(j)} Z_{(k)} Z_{(l)} \nonumber \\
&&+\text{ terms of the type } \;d_{ij...m}Z_{(j)} Z_{(k)} ...Z_{(m)} \text{ ( $\leq 6^{th}$ degree)} \nonumber \\
&& + e_{ijk}r Z_{(j)}^{\prime}Z_{(k)} + f_{ijkl}r Z_{(j)}^{\prime}Z_{(k)}Z_{(l)} \,, \label{EqZ}
\end{align}
where $i=\{\Yb,\,V,\,F,\,\Lb\,,\pb\}$. Now, if we consider, as we discussed in Sec. \ref{PT}, this system in successive nonlinear iterations, for each approximation order $n$, it will look as follows
\begin{align}
&&r Z^{\prime}_{(i)n}+ c_{ij}Z_{(j)n} = \sum_{n_1+n_2=n} a_{ijk}Z_{(j)n_1} Z_{(k)n_2} \nonumber \\
&&+\sum_{n_1+n_2+n_3=n} b_{ijkl}Z_{(j)n_1} Z_{(k)n_2} Z_{(l)n_3} \nonumber \\
&&+\text{ terms of the type } \nonumber \\ 
&&\sum_{n_1+n_2+..+n_t=n} \;d_{ij...m}Z_{(j)n_1} Z_{(k)n_2} ...Z_{(m)n_t} \text{ ( $t\leq 6$ )} \nonumber \\
&& + \sum_{n_1+n_2=n} e_{ijk}r Z_{(j)n_1}^{\prime}Z_{(k)n_2} \nonumber \\
&&+ \sum_{n_1+n_2+n_3=n} f_{ijkl}r Z_{(j)n_1}^{\prime}Z_{(k)n_2}Z_{(l)n_3} \,. \label{EqZn}
\end{align}

Let us first consider the second order of expansion in nonlinearity. At the second order we will have a system of the type
\begin{align}
&&r Z^{\prime}_{(i)2}+ c_{ij}Z_{(j)2} = a_{ijk}Z_{(j)1} Z_{(k)1}  \nonumber \\
&& + e_{ijk}(r Z_{(j)1}^{\prime})Z_{(k)1}  \,.  \label{EqZ2}
\end{align}
Taking into account \eqref{FeatureZ1} we conclude that the right hand side of Eq. \eqref{EqZ2} reads
\bea
&&r Z^{\prime}_{(i)2}+ c_{ij}Z_{(j)2}  \sim \left(\frac{m_1}{r}\&\frac{ C }{r^3}\right) \times \left(\frac{m_1}{r}\&\frac{ C }{r^3}\right)\nonumber \\
&&  \sim \left( \frac{m_1}{r} \right)^2 \& \frac{m_1}{r}\frac{ C }{r^3}\&\left( \frac{ C }{r^3} \right)^2  \,.  \label{EqZ22}
\eea
A decaying solution of a system of equations of the type \eqref{EqZ22}, with a right-hand side being a polynomial in $m_1/r$ and $C/r^3$, 
is given by polynomials in $m_1/r$ and $C/r^3$ of the same power as on the right-hand side. In other words, 
\be
Z_{(i)2} \sim \left( \frac{m_1}{r} \right)^2 \& \frac{m_1}{r}\frac{ C }{r^3}\&\left( \frac{ C }{r^3} \right)^2   \label{FeatureZ2}
\ee
which means more explicitly
\be
Z_{(i)2} = v_i\left( \frac{m_1}{r} \right)^2 + \overline{v_i}\frac{m_1}{r}\frac{ C }{r^3}+ \overline{\overline{v_i}}\left( \frac{ C }{r^3} \right)^2  
\ee
with some numerical coefficients $v_i, \; \overline{v_i}, \; \overline{\overline{v_i}}$.

Indeed, solving the system \eqref{Sysnn1}--\eqref{EqLnn} outside the source in the second order in nonlinearity, we obtain the following explicit exterior solution
\begin{eqnarray}
Z_{(m0)2}&=-&\frac{2m_1}{3r}(1+\eta)\left(  -\frac{m_1}{r} + \frac{9 C }{r^3}  \right)  \nonumber \\
Z_{(mk)2}&=&-\frac{m_1}{2r}(1+\eta)\left(\frac{m_1}{r} - \frac{8C}{r^3} \right) \nonumber \\
Z_{(V)2}&=&\frac{2m_1}{3r}(1+\eta)\left(  -\frac{m_1}{2r} + \frac{3 C }{r^3}  \right)  \nonumber \\
Z_{(\Yb)2}&=&-\frac{1}{6}(1+\eta)\left(\frac{m_1}{r} \right)^2 \nonumber \\
Z_{(\pb)2}&=&\frac{3m_1 C (3\eta+4)}{r^4} - \frac{m_1^2(14+15\eta)}{6r^2} \nonumber \\
Z_{(F)2} &=& -\frac{4m_1 C \eta(1+\eta)}{r^4} + \frac{m_1^2}{9r^2}(18+44\eta+25\eta^2) \nonumber \\
Z_{(\Lb)2} &=& \frac{m_1 C \eta(1+\eta)}{r^4} + \frac{m_1^2}{18r^2}(27+44\eta+19\eta^2) \,. \nonumber 
\end{eqnarray}
Since all the $Z_{(i)2}$ are polynomial in $m_1/r$ and $C/r^3$, we can write for them the same relation as \eqref{FeatureZ1}:
\be
r Z_{(i)2}^{\prime}=\left( \frac{m_1}{r} \right)^2 \& \frac{m_1}{r}\frac{ C }{r^3}\&\left( \frac{ C }{r^3} \right)^2 \,. \label{FeatureZ2d}
\ee
Note that the above (decaying) inhomogeneous solutions are uniquely determined modulo the addition of $O(\varepsilon^2)$
 homogeneous solutions of the linearized (first-order) equations. The latter additions can be simply absorbed by making some renormalizations of the parameters $m_1$ and $ C $ entering the first-order solution (which were defined as  first-order variables) of the type 
 $\varepsilon m_1 \to \varepsilon m_1+ \varepsilon^2 m_2$,
  $\varepsilon C \to \varepsilon C + \varepsilon^2 C_2$. Therefore, when describing the structure of the general exterior solution
  it is sufficient to construct at each order of nonlinearity expansion a (decaying) inhomogeneous solution of the type of $Z_{(i)2}$
  above, with the understanding that the basic parameters $m_1$ and $ C $ entering the first-order solution might absorb 
  renormalization corrections of all higher orders.

Then, continuing in the same way, we can write equations \eqref{EqZn} for $n=3$:
\begin{align}
&&r Z^{\prime}_{(i)3}+ c_{ij}Z_{(j)3} = a_{ijk}Z_{(j)1} Z_{(k)2} + b_{ijkl}Z_{(j)1} Z_{(k)1} Z_{(l)1} \nonumber \\
&& + e_{ijk}r Z_{(j)1}^{\prime}Z_{(k)2} + e_{ijk}r Z_{(j)2}^{\prime}Z_{(k)1} \nonumber \\
&&+ f_{ijkl}r Z_{(j)1}^{\prime}Z_{(k)1}Z_{(l)1}  \,. \label{EqZ3}
\end{align}
The right-hand side of \eqref{EqZ3}, taking into account \eqref{FeatureZ2} and \eqref{FeatureZ2d}, will have the following form
\be
\left(\frac{m_1}{r}\&\frac{ C }{r^3}\right) \times \left(\frac{m_1}{r}\&\frac{ C }{r^3}\right) \times \left(\frac{m_1}{r}\&\frac{ C }{r^3}\right)
\ee
The solution of \eqref{EqZ3} will then look as
\bea
&&Z_{(i)3} \sim \left(\frac{m_1}{r}\&\frac{ C }{r^3}\right) \times \left(\frac{m_1}{r}\&\frac{ C }{r^3}\right) \times\left(\frac{m_1}{r}\&\frac{ C }{r^3}\right) \nonumber \\
&& \sim \l\frac{m_1}{r} \r^3 \& \l \frac{m_1}{r} \r^2 \l \frac{C}{r^3} \r \& \l \frac{m_1}{r} \r \l \frac{C}{r^3} \r^2 \& \l \frac{C}{r^3} \r^3 \,. \nonumber \\
\eea
modulo some {\it  logarithms} that will be discussed later. 

In the higher orders of perturbation theory we will have, by induction, a nonlinearity expansion
$ Z_{(i)} = \sum_{n=1}^{n=\infty} Z_{(i)n}$ with a $n^{\rm th}$ order that we can symbolically write as
\be
Z_{(i)n} \sim  \l \frac{m_1}{r}\& \frac{ C }{r^3}\r^n \;, \label{SolZn}
\ee
 where it understood that $(X\&Y)^n$ denotes a homogeneous polynomial of order $n$ in $X$ and $Y$,
 i.e. $X^n \& X^{n-1} Y \& \cdots \& Y^n$.

In the reasonings  used in the derivation of the formula \eqref{SolZn} we did not take into account the presence of {\it logarithmic terms} $\log{r}$. Such terms  appear each time the source term of the equation for $Z_{(i)n}$  contains the power of $1/r$ which matches a
(decaying) homogeneous solution of this equation.

Let us look at these logarithmic terms more precisely.
We write down the two differential equations \eqref{Sysnn1} and \eqref{Sysnn2} in dimensionless form and for $n\geq 3$. They read as follows 
\begin{eqnarray}
&& r^2{V^{m0}_n}^{\prime} +2rV^{m0}_n\nonumber \\
&&= N^{m0}_{n\;\k=0} r^2 \equiv \sigma^{m0}_n \,, \label{dimlessVm0}  \\
&& r^2{\pb_n}^{\prime\prime}+2r{\pb_n}^{\prime} - 6\pb_n \nonumber \\
&&= 6rV^{m0}_n   +r^2{\widehat{N}^{\pb}}_{n\;\k=0} \equiv \sigma^{\pi}_n   \,. \label{dimlessp}
\end{eqnarray}
Since Eqs. \eqref{dimlessVm0}, \eqref{dimlessp} can be derived from the system \eqref{InEqBegin}-\eqref{InEqEnd} written in the form \eqref{EqZ}, the right-hand sides of \eqref{dimlessVm0}, \eqref{dimlessp} have the form 
 $$\sigma^{m0}_n \sim \sigma^{\pi}_n \sim \l\frac{m_1}{r}\&\frac{ C }{r^3}\r^n \,. $$
So Eqs. \eqref{dimlessVm0}--\eqref{dimlessp} read
\begin{eqnarray}
&& r^2{V^{m0}_n}^{\prime} +2rV^{m0}_n \sim \sum_{s }\frac{1}{r^s}  \,,\label{dimlessVm0r} \\
&& r^2{\pb_n}^{\prime\prime}+2r{\pb_n}^{\prime} - 6\pb_n \sim \sum_{p}\frac{1}{r^p}  \,,\;\; s,\,p>0 \,. \label{dimlesspr}
\end{eqnarray}

Logarithms will enter in the solution of Eq. \eqref{dimlesspr} when $p=-2$ or $p=3$. The first case  ($p=-2$) never occurs because we
constructed a decaying solution. The second case ($p=3$) is realized in the third order in perturbation theory because of the term 
$$ \frac{m_1^3}{r^3} \,. $$
 Indeed, the source term for $\pb_3$ is
\begin{eqnarray}
{\sigma^{\pi}_3\;}&=&\frac{2m_1^3(1+\eta)(37+59\eta)}{9r^3}  \nonumber \\
&+& \frac{m_1^2 C (234+383\eta+163\eta^2)}{r^5} + \frac{162m_1 C ^2(1+\eta)^2}{r^7} \,. \nonumber \\ \label{Source3}
\end{eqnarray}
Concerning Eq. \eqref{dimlessVm0r}, logarithms could only appear there for $s=1$, which is impossible 
because the nonlinearity order $n>1$. 

To further clarify the situation with logarithms, let us consider the equation for $\pb$ in the $n$th order  of perturbation theory and try to
construct  a solution using the Green's function $G_0(r,r')=\lim_{\k\to0} G_\k(r,r')$ . According to Eqs. \eqref{GreenFunc}, \eqref{Wronsk} and \eqref{dimlessp} we can write the following exterior solution for $\pb_n$ 
\begin{eqnarray}
\pb_n^{(r>R_S)} = &&X_>(r)\int_0^{R_S}X_<(\hr){\sigma^{\pi}_n\;} (\hr)d\hr  \nonumber \\
&& + X_>(r)\int_{R_S}^{r}X_<(\hr){\sigma^{\pi}_n\;} (\hr)d\hr \nonumber \\
&& + X_<(r)\int_r^{\infty}X_>(\hr){\sigma^{\pi}_n\;} (\hr)d\hr \nonumber \\
 =&& -\frac{1}{5}\frac{1}{r^3}\int_{0}^{R_S}\hr^2 {\sigma^{\pi}_n\;} (\hr) d\hr \label{Int1} \\
&& -\frac{1}{5}\frac{1}{r^3}\int_{R_S}^{r}\hr^2{\sigma^{\pi}_n\;} (\hr)d\hr \label{Int2} \\
&&  -\frac{1}{5}r^2\int_{r}^{\infty}\frac{1}{\hr^3}{\sigma^{\pi}_n\;} (\hr) d\hr  \,.\label{Int3}
\end{eqnarray}
When  $${\sigma^{\pi \, (r>R_S)}_n\;}  \sim \frac{1}{r^3}$$
the integral \eqref{Int2} gives 
\be
\frac{1}{r^3}\int_{R_S}^{r}\hr^2{\sigma^{\pi}_n\;} (\hr)d\hr \propto \frac{1}{r^3}\log{\frac{r}{R_S}}\,.
\ee
The other terms do not generate logarithms. The term  \eqref{Int1} generates a $O(1/r^3)$ contribution in the exterior.
And one easily checks that the source terms  ${\sigma^{\pi \, (r>R_S)}_n\;} \sim r^{-p}$ with $p \neq 3$ generate corresponding
exterior solutions through the integrals \eqref{Int2} and  \eqref{Int3} which are again $\sim r^{-p}$.
Moreover, the infinite-range integral \eqref{Int3} is always convergent. 

So, a logarithm appears at third order, 
\be
\pb_3^{(r>R_S)} \sim \frac{m_1C^2}{r^7} \& \frac{m_1^2 C }{r^5} \& \frac{m_1^3}{r^3}(1 \& \log{\frac{r}{R_S}})\,,
\ee
where (in view of the integral \eqref{Int2}) it is natural to take $R_S$ as the scale appearing  in the logarithm. 
Once appeared, this logarithm will cause the appearance of logarithms in the next orders in all the variables. For example, the fourth-order source in equation \eqref{dimlessp} will contain a term of the type
\begin{eqnarray*}
{\sigma^{\pi}_4\;}=&...& + \left(  \frac{m_1}{r} \& \frac{ C }{r^3} \right)  \left[  \frac{m_1 C ^2}{r^7} \& \frac{m_1^2 C }{r^5} \right. \\
&\&& \left. \frac{m_1^3}{r^3}\left( 1 \& \log{\frac{r}{R_S}}  \right) \right]\,.
\end{eqnarray*}
The corresponding solution will have the same structure
\begin{eqnarray*}
{\pb}_4 \;=&...& + \left(  \frac{m_1}{r} \& \frac{ C }{r^3} \right)  \left[  \frac{m_1 C ^2}{r^7} \& \frac{m_1^2 C }{r^5} \right. \\
&\&& \left. \frac{m_1^3}{r^3}\left( 1 \& \log{\frac{r}{R_S}}  \right) \right]\,.
\end{eqnarray*}
It is, indeed, important to note that, beyond the third order of nonlinearity, the source terms will all decay strictly faster than $1/r^3$
(modulo logarithmic factors), so that the power of the logarithm appearing in the source term will remain the same in the
corresponding solution. The only way the power of the logarithm will increase is then through nonlinear combinations
of the previously generated $\log (r/R_S)/r^p$ contributions.

For instance, $\sigma_{(i)\, 6}$ will contain, among  other terms, also $\pb_3\pb_3$ which will give rise to a $\log^2(r/R_S)$
factor.  This squared logarithm will remain so until one reaches the ninth iteration level where  $\sigma_{(i)\, 9}$
 will contain  $\pb_3^3$, and therefore  $Z_{(i)\, 9}$ will involve a $\log^3(r/R_S)$ factor, etc.

We can summarize this structure by writing that, in its dimensionless form \eqref{Zdef}, the all-order exterior
 solution for $\k=0$ has the following form
\be
Z_{(i)}=\sum_{n}\left[\frac{m^3}{r^3}\log{\frac{r}{R_S}}\&\l\frac{m}{r}\&\frac{\hC}{r^3}\r^3\right]^{\left[\frac{n}{3}\right]} \left(  \frac{m}{r}\&\frac{\hC}{r^3}  \right)^{n-3\left[\frac{n}{3}\right]} \,, \label{FullSol}
\ee
where $\left[\frac{n}{3}\right]$ denotes the integer part of $n/3$. In addition, we have denoted $m\equiv m_1+\delta m_1$ and $\hC \equiv C+ \delta  C $, where $\delta m_1$ and $\delta C$ are some renormalizations of the first-order parameters $m_1$ and $  C $ 
(as discussed above).

\section{Solution in the region $\k r\gg 1$}

Up to now we have focussed on the structure of the exterior solution in the region $\k r \ll1$ (when considering a very small $\k$).
The structure we found is expected to be  physically accurate as long as $ \k r \lesssim 1$. 
Let us now discuss the expected structure of the exterior solution in the complementary region   $ \k r \gtrsim 1$,
mathematically described by considering the limit $\k r \gg 1$. We have already shown in our previous paper \cite{Damour:2019oru} 
that torsion components are exponentially decaying $\sim e^{- \k r}/r^n$. This fact, together with the form of the linear solution \eqref{pblin}--\eqref{pbadd}, and the form of the second-order solution computed in our previous paper (see \cite{Damour:2019oru} Eq. (8.9)--(8.18)), suggests that the solution of our field equations in the domain $\k r \gg 1$ has the following approximate form:
\be
Z^{\k r \gg 1}_{(i)}= Z^{\rm powerlaw}_{(i)} + \sum_{p \geq1,n} C_{(i)np}\frac{e^{-p\k r}}{r^n} \,, \label{ZLaw}
\ee
were $C_{(i)np}$ are some coefficients  which are {\it regular as $\k\to0$}. Furthermore, the exponential series in \eqref{ZLaw} starts as 
the first-order solution we found in  Sec.~\ref{Linear}: for instance, according to \eqref{Vmklin}, $$ C_{(mk)01}=\k C_{(mk)11} = -\l \frac{2m_1}{3}+C\k^2 \r  \,.$$

Let us clarify the structure of the power-law contribution $Z^{\rm powerlaw}_{(i)}$ in the above expression. To this end, let us recall the 
vacuum ($T_{ij}=0$) vierbein field equation taken from Eq. (3.2) in Ref. \cite{Nikiforova:2018pdk}. There, that field equation was general, and was
also written for a more general model than the torsion bigravity one. When restricting to the torsion bigravity model,  
and considering a symmetric $F_{ij}$ (as appropriate to our static spherically symmetric case), this equation reads ${{\cal G}}_{ij} =0$
where
\begin{align} \label{Geq}
{{\cal G}}_{ij} \equiv & 
 c_F \left( F_{ij} - \frac{1}{2} \eta_{ij}F  \right)+c_R \left( R_{ij} - \frac{1}{2} \eta_{ij}R  \right) \nonumber  \\ 
&  + c_{F^2} \left[ F_{ki} F_{kj} + F_{kl} F_{kilj} - \frac{2}{3}F\,F_{ij}  \right.
\nonumber\\
&\left.  - \frac{1}{2} \eta_{ij}\l F_{kl}F_{kl} - \frac{1}{3}F^2 \r  \right]= 0 \;.
\end{align}
At  large distances $\k r \gg 1$, where the torsion vanishes, we have $F_{ijkl}=R_{ijkl}$, and $F_{ij}=R_{ij}$
so that Eq. \eqref{Geq} takes the following form
\begin{align} \label{Geq2}
 &(c_R+c_F) \left( R_{ij} - \frac{1}{2} \eta_{ij}R  \right) \nonumber  \\ 
&  + \frac{\eta \lambda}{\k^2} \left[ R_{ki} R_{kj} + R_{kl} R_{kilj} - \frac{2}{3}R\,R_{ij}  \right.
\nonumber\\
&\left.  - \frac{1}{2} \eta_{ij}\l R_{kl}R_{kl} - \frac{1}{3}R^2 \r  \right]= 0 \;.
\end{align}
Here, we used the shorthand notation that frame indices contracted by the Minkowski metric are all formally written as covariant indices.

As the Riemannian curvature tensor $R_{ijkl}$ is, by assumption,  decaying with $1/r$, the terms quadratic in $R_{ij}$,
or bilinear in $R_{ij}$ and $R_{ijkl}$ are decaying at large distances in a faster way than the linear-in-Ricci term
on the first line. This is easily seen to imply that the only $r$-decaying solution of Eq. \eqref{Geq2} is $R_{ij}=0$.
By Birkhoff theorem the only  spherically symmetric (zero torsion) solution of $R_{ij}=0$  is the Schwarzshild metric. 
Thus, we conclude that the power-law contribution in \eqref{ZLaw} is described by a Schwarzschild solution
\bea
L_S\equiv e^{\Lambda_S}&=&\frac{1}{\sqrt{1-2\frac{m_S}{r}}}  \label{LFar} \\
\Lb_S\equiv e^{\Lambda_S}-1&=&\frac{1}{\sqrt{1-2\frac{m_S}{r}}}-1  \label{LFar} \\
F_S \equiv \Phi^{\prime}_S&=&\frac{m_S}{r^2(1-2\frac{m_S}{r})}  \\
V_S&=&\frac{F_S}{L_S}=\frac{m_S}{r^2\sqrt{1-2\frac{m_S}{r}}}  \\
W_S&=&-\frac{\sqrt{1-2\frac{m_S}{r}}}{r} \,, \label{WFar}
\eea
where
$$
ds^2=-e^{2\Phi_S}dt^2 + e^{2\Lambda_S}dr^2 + r^2\left( d\theta^2+\sin^2\theta\, d\phi^2 \right) 
$$
is a Schwarzschild metric (of mass $m_S$), and the functions $V_S$ and $W_S$ are determined from the condition $K_{ijk}=0$ (see Eq. \eqref{contorsion}). In the dimensionless form $Z_{(i)}$ the expressions \eqref{LFar}--\eqref{WFar} read
\bea
Z^{\rm Schw}_{(\Lb)}&=&\frac{1}{\sqrt{1-2\frac{m_S}{r}}}-1 \label{ZSL} \\
Z^{\rm Schw}_{(F)}&=&\frac{m_S}{r(1-2\frac{m_S}{r})} \\
Z^{\rm Schw}_{(V)}&=&\frac{Z^{\rm Schw}_{(F)}}{1+ Z^{Schw}_{(\Lb)}}=\frac{m_S}{r\sqrt{1-2\frac{m_S}{r}}} \\
Z^{\rm Schw}_{(\Yb)}&=&\frac{3m_S-r}{r\sqrt{1-\frac{2m_S}{r}}}+1 \,, \label{ZSYb}
\eea
where $Z^{\rm Schw}_{(i)}$ are defined by \eqref{Zdef}.
Then, according to Eq. \eqref{defPb}, $\pb_S$ (the value of $\pb$ corresponding to  a Schwarzschild solution) reads
\be
\pb_S=Z_{(\pb)}^{\rm Schw}=-\frac{6m_S}{\k^2 r^3} \label{pbFar} 
\ee
(compare to \eqref{pbFirst}). The factor $\k^2$ in the denominator of Eq. \eqref{pbFar} is not jeopardizing for the small $\k$ limit, because 
we are now considering the $\k r \gg 1$ region, so that
$$ \pb_S=-\frac{6m_S}{\k^2 r^3}=-\frac{6m_S}{r}\frac{1}{\k^2r^2} \ll \frac{m_S}{r}\,.$$ 

Finally, the solution at large distances has the following form
\be 
Z_{(i)}^{\k r\gg 1} = Z^{\rm Schw}_{(i)} + \sum_{p\geq 1,n} C_{(i)np}\frac{e^{-p\k r}}{r^n} \,  \label{ZFarEx}
\ee
with $Z^{\rm Schw}_{(i)}$ given by \eqref{ZSL}--\eqref{pbFar} and $C_{(i)np}$ regular as $\k\to0$.
Let us note in passing that one can easily set up a formal perturbation formalism for computing the exponentially decaying piece
in Eq. \eqref{ZFarEx}. Indeed, in the region $\k r \gtrsim 1$ one can describe the looked-for solution as a perturbation
(in torsion bigravity theory) of a zero-torsion Schwarzschild solution (which is an exact background solution of torsion bigravity).
The first steps for studying torsion-bigravity perturbation theory around general Einstein backgrounds were set up
in Refs. \cite{Nikiforova:2009qr,Deffayet:2011uk}.

Above we have described in some detail the structure of solutions of torsion bigravity (in the small-mass regime)  in
the two different regions $\k R_S < \k r \lesssim 1$ and $\k r \gtrsim 1$
by means of two different expansions, see Eqs. \eqref{FullSol} and \eqref{ZFarEx}.
The use of the general perturbation theory set up in Sec. V above yields, in principle, a way to explicitly describe the solution
in the full domain $0\leq \k r <\infty$. As we have used in the construction of Eq. \eqref{FullSol} the $\k \to 0$ limit
of the general Green's function $G_\k(r,r')$ used in our general perturbation theory, and found that it led to constructing a
physically meaningful solution, we expect that the so-constructed general expansion would interpolate
between the expansions Eqs. \eqref{FullSol} and \eqref{ZFarEx}, valid  in the two limiting regions $ \k R_S <\k r \ll 1$ and $\k r \gg 1$.

\section{Conclusions}

Working within a static spherically symmetric ansatz, we provided detailed evidence that there exists a regular finite massless limit, 
$\k = m_2 \to 0$, in torsion bigravity, i.e. that, contrary to all previously studied (nonlinear Fierz-Pauli-type) models, there appear no 
inverse powers of the mass $\k = m_2$ of the massive spin-2 excitation when solving the theory by a weak-field perturbation theory. 
In other words, there appear no Vainshteinlike scale limiting the domain of validity of perturbation theory.
This makes torsion bigravity dramatically  different from the standard theories of massive gravity and bigravity. 

The root of this basic difference is that torsion bigravity is a completely new type of theory in which the massive spin-2 excitation
has  a geometric origin, and is contained in the torsion of an independent affine connection, rather than in a second
metric tensor (as in bimetric gravity). Taking into account the fact that, in our previous work \cite{Damour:2019oru}, we have found
 that the number of degrees of freedom (within a static spherically symmetric ansatz) in this theory is the same as in {\it ghost-free} bigravity, it is clear that torsion bigravity deserves further study. 

In the previous work \cite{Damour:2019oru} we discussed the phenomenology of  torsion bigravity based on the result of  second-order 
perturbation theory, assuming that the latter one gives a sufficiently accurate description of the deviations from GR. 
Indeed, we had left open the possibility that some  Vainshteinlike scale, appearing at the cubic order of perturbation theory,
would limit the validity of perturbation theory, and would invalidate any phenomenological conclusion drawn from second-order
perturbative solutions. As we have found here
the absence (at all perturbative orders) of any Vainshteinlike scale limiting the validity of perturbation theory, we can trust the conclusions 
reached in our previous work.  In \cite{Damour:2019oru} we derived (in the regime where $\k^{-1}$  is much larger that the length 
scales that are being experimentally probed) a strong phenomenological
constraint on the parameter $\eta$ (the ratio of the massive and the massless couplings to matter, see \eqref{cF}), namely 
\be 
\eta \lesssim 10^{-5} \,. \label{Limit} 
\ee  
The reason for such a strong limit is that the vDVZ discontinuity is present in torsion bigravity. As we have now proven that there are no Vainshteinlike radius, and therefore no possible Vainshtein screening, it is necessary, in view of the very accurate
confirmation of GR on large scales, to constrain the strength of the coupling to matter of the massive spin-2 perturbation excitation.
Note, in this respect, that the limit \eqref{Limit} can be  weakened if we would be considering the case of smaller ranges  $\k^{-1}$
for the massive excitation.

Even when considering large ranges $\k^{-1}$, the limit \eqref{Limit} does not necessarily mean that the theory always behaves 
very close to GR. In particular, though the Schwarzschild (and Kerr) black holes are exact solutions of torsion bigravity, there might
exist black holes endowed with torsion hair in torsion bigravity. We leave this interesting issue to future study.

\section*{Acknowledgments}
I thank Thibault Damour and C\'edric Deffayet for numerous informative discussions.


\end{document}